\documentclass[journal]{IEEEtran}
\usepackage{amsmath,amsfonts}
\usepackage{algorithmic}
\usepackage{algorithm}
\usepackage{array}
\usepackage[caption=false,font=normalsize,labelfont=sf,textfont=sf]{subfig}
\usepackage{textcomp}
\usepackage{lettrine}
\usepackage{stfloats}
\usepackage{url}
\usepackage{verbatim}
\usepackage{graphicx}
\usepackage[utf8]{inputenc}
\usepackage{cite}
\usepackage{enumitem}
\usepackage{booktabs}
\usepackage{tabularx}
\usepackage{multirow}
\usepackage{array}
\usepackage{booktabs}
\usepackage{makecell}
\newcolumntype{Y}{>{\centering\arraybackslash}X}
\usepackage{subfig}

\usepackage{siunitx}
\usepackage[colorlinks,
            linkcolor=blue,
            citecolor=blue,
            urlcolor=blue]{hyperref}

\def\BibTeX{{\rm B\kern-.05em{\sc i\kern-.025em b}\kern-.08em
    T\kern-.1667em\lower.7ex\hbox{E}\kern-.125emX}}

\begin{document}

\title{\fontsize{24}{28}\selectfont
On Spatial Degree-of-Freedom Analysis of Near-Field Multipath Channels for Ultra-massive MIMO Systems}

\author{
    {Zhiqiang Yuan, \textit{Member, IEEE}, Hui Lou, Wei Fan, \textit{Senior Member, IEEE}, \\Jianhua Zhang, \textit{Fellow, IEEE}, and Henk Wymeersch, \textit{Fellow, IEEE}}
    \thanks{This work was supported by HORIZON-MSCA NEDS-6G Project under Grant 101209281. (Corresponding author: Wei Fan.)}%
    \thanks{Zhiqiang Yuan and Henk Wymeersch are with the Department of Electrical
Engineering, Chalmers University of Technology, 41296 Gothenburg, Sweden (e-mail: \{yuanzhiq, henkw\}@chalmers.se).}%
    \thanks{Hui Lou and Wei Fan are with National Mobile Communications Research Laboratory, School of Information Science and Engineering, Southeast University, Nanjing 210096, China (e-mail: weifan@seu.edu.cn).}%
    \thanks{Jianhua Zhang is with the State Key Lab of Networking and Switching Technology, Beijing University of Posts and Telecommunications, Beijing 100876, China (e-mail: jhzhang@bupt.edu.cn).}%
}

\maketitle

\begin{abstract}
The transition to near-field (NF) communications in ultra-massive multiple-input multiple-output (UM-MIMO) systems fundamentally alters the spatial degrees of freedom (DoF) of wireless channels. While the NF DoF of line-of-sight (LoS) transmission channels is well-characterized in the literature, the DoF in NF multipath scenarios remains underexplored. This paper investigates the spatial DoF of NF UM-MIMO channels under practical multipath conditions. A generic DoF metric is derived by modeling multipath propagation and analyzing the resulting eigenvalue distribution based on the Green’s function representation of the channel. The DoF contribution of each path is determined by the product of the effective electrical aperture and the subtended solid angle, and the total DoF is obtained through the effective union of spatially resolvable path contributions. A mapping between the eigenvalue distribution and multipath powers is further established. Numerical simulations and real-world NF channel measurements at 28-30 GHz with 720 array elements are conducted for validation in both LoS multipath and non-LoS scenarios. The results show that multipath propagation can significantly increase the spatial DoF and that the proposed metric accurately predicts the DoF of practical NF channels. The proposed framework provides a practical tool for DoF prediction and supports capacity analysis and spatial multiplexing design in future NF UM-MIMO systems.
\end{abstract}   

\begin{IEEEkeywords}
Spatial DoF, NF, UM-MIMO channel measurement, multipath.
\end{IEEEkeywords}

\section{Introduction}

Massive and ultra-massive multiple-input multiple-output (UM-MIMO) are widely recognized as key enabling technologies for 6G communications \cite{tataria20216g}. By employing a very large number of antennas (ranging from tens to thousands), especially paired with millimeter-wave (mmWave) frequencies for compact deployments, UM-MIMO not only inherits the fundamental advantages of conventional MIMO systems but also significantly enhances spectral efficiency, energy efficiency, and link reliability \cite{molisch2006comprehensive,saad2019vision}. However, the adoption of large array apertures and high carrier frequencies significantly enlarges the Rayleigh distance, shifting wireless propagation from the conventional far-field (FF) region to the radiating near-field (NF) region \cite{lu2021communicating}. This physical shift invalidates the traditional FF plane-wave assumption from the electromagnetic perspective \cite{yuan2022spatial}, necessitating a rigorous re-evaluation of fundamental limits in NF spherical-wave propagation for UM-MIMO systems \cite{giordani2020toward}. 


One of the most important fundamental limits is the spatial degrees of freedom (DoF), an intrinsic property of wireless channels that defines the upper-bound for the number of independent waveforms available for data transmission \cite{migliore2006role,pizzo2022nyquist,poon2005degrees}. It serves as a key indicator for practical array deployment and for transmission strategy selection (e.g., spatial multiplexing or beamforming) \cite{haneda201360,lyu2024large}. Extensive studies have characterized the spatial DoF of multi-antenna channels in FF conditions from various perspectives, e.g., leveraging optic diffraction theory \cite{thaning2003limits}, signal-space analysis \cite{pizzo2022nyquist}, and electromagnetics \cite{bucci1989degrees}. These analyses show that the DoF of an FF channel depends on the minimum of the number of transmitter (Tx) and receiver (Rx) array elements and the number of resolvable propagation paths in the channel. In particular, for a line-of-sight (LoS) transmission channel (i.e., only the LoS path exists) in the FF, the DoF is equal to 1. However, this DoF guideline for FF cases cannot directly extend to NF cases due to the introduced spherical-wave propagation \cite{wang2025analytical,kosasih2025spatial}. A well-known example is that the DoF of a LoS-transmission channel in the NF is demonstrated to (even largely) exceed 1 \cite{lu2021communicating}. In principle, spherical waves in the NF region introduce distance-dependent phase variations across the array aperture, creating additional spatially distinguishable modes and thereby increasing the achievable DoF. Owing to this potential DoF enhancement, the characterization of NF spatial DoF has attracted growing attention in recent UM-MIMO studies.

The spatial DoF of NF LoS-transmission channels was analyzed in \cite{miller2000communicating,poon2005degrees} through eigenvalue decomposition of Green’s function, where the DoF was derived as the product of the Tx array aperture and the channel solid angle subtended by the Rx array projection along the LoS direction. An alternative derivation based on Landau’s eigenvalue theorem was presented in \cite{pizzo2022landau}, confirming the validity of this DoF metric for LoS-transmission cases, and simulations under various array apertures and carrier frequencies were performed for validation. In \cite{do2022line}, the metric was further extended to cases of LoS MIMO transmission via reconfigurable intelligent surfaces (RISs), showing that the DoF scales proportionally with the RIS aperture. Based on this metric, several studies have investigated NF communication characteristics, including DoF enhancement compared to the FF regime \cite{kosasih2025spatial}, NF–FF boundary analysis \cite{10819602}, and beamforming design for maximum multiplexing gain \cite{ouyang2023near}. However, these works mainly focus on LoS transmission, whereas practical NF channels exhibit multipath propagation in both LoS and non-LoS (NLoS) scenarios \cite{lyu2024large}. In \cite{pizzo2022nyquist}, Nyquist sampling under arbitrary NF scattering conditions was investigated, indicating a DoF increase in multipath environments compared to the LoS-transmission case. Similarly, \cite{wan2024near} studied NF channel modeling from an electromagnetic information perspective and evaluated the corresponding DoF through multipath channel simulations. Nevertheless, a clear metric for DoF calculation in NF multipath channels remains lacking.
In summary, the state-of-the-art metric for NF DoF calculation mainly focuses on LoS transmission, while a generic metric applicable to practical NF multipath channels is still missing. Moreover, experimental results of NF DoF analysis are rarely reported in the literature, leaving the existing metric insufficiently verified and generic and effective DoF metrics for NF multipath channels largely unexplored.

To fill those gaps, we analyze the spatial DoF for UM-MIMO systems in this paper. We first reveal the limitations of the existing DoF calculation metric in practical UM-MIMO deployments, and then propose a novel metric for generic NF multipath scenarios. Moreover, real-world channel measurements are performed, validating the effectiveness of the proposed metric. The main contributions of this paper are summarized as follows:
\begin{itemize}
    \item Through theoretical analysis and introduction of a measured multipath channel, we reveal that the existing NF DoF metric, derived under the LoS transmission assumption, cannot accurately characterize practical NF channels with multipath propagation. The measurement results show that the spatial DoF in NF multipath channels can be significantly larger than that predicted by the existing metric, highlighting the important contribution of multipath components to the spatial DoF.
    \item A generic DoF metric is derived for practical NF multipath channels. The DoF contribution of each multipath is derived as the product of the effective electrical aperture and the subtended solid angle, and the total channel DoF is characterized through the effective union of spatially resolvable path contributions. In addition, the relationship between the eigenvalues and path powers is revealed, providing a physical interpretation of the spatial modes and insights to guide the metric implementation under different power conditions.
    \item The proposed metric is validated through numerical simulations and real-world NF channel measurements at 28-30 GHz using a 720-element virtual array, covering both LoS multipath and NLoS scenarios. Results show that the DoF predicted by the proposed metric agrees well with the eigenvalue distributions of the measured channels. A significant increase in spatial DoF due to multipaths is observed and accurately captured by the proposed metric, further verifying the effectiveness of the analysis.
\end{itemize}

The remainder of this paper is organized as follows: Section~\ref{sec:system_model} presents the NF UM-MIMO system model and analyzes the limitations of the existing DoF metric in multipath scenarios. In Section~\ref{third}, we derive the generic DoF metric for practical NF multipath scenarios and detail the metric implementation. Section~\ref{fourth} provides simulation results to demonstrate the proposed metric, and Section~\ref{fifth} presents the channel measurements for experimental validation. Section~\ref{sixth} concludes the paper.

Notation: Scalars are denoted by regular font (e.g., $a$), vectors by bold lowercase font (e.g., $\mathbf{a}$), and matrices by bold uppercase font (e.g., $\mathbf{A}$). The superscripts $(\cdot)^*$ and $(\cdot)^\mathrm{H}$ represent the complex conjugate and conjugate transpose operators, respectively. $\|\cdot\|$ denotes the Euclidean norm, while $|\cdot|$ represents the absolute value of a scalar. $\lfloor \cdot \rfloor$ denotes the floor function (rounding down to the nearest integer), and $\text{vec}\{\cdot\}$ represents the vectorization operator, which stacks the columns of a matrix into a single column vector. Furthermore, $\mathbb{C}^{M \times N}$ denotes the space of $M \times N$ complex matrices, and $\cup$ denotes the set union operator.

\begin{figure}[!t]
\centering
\includegraphics[width=1\linewidth]{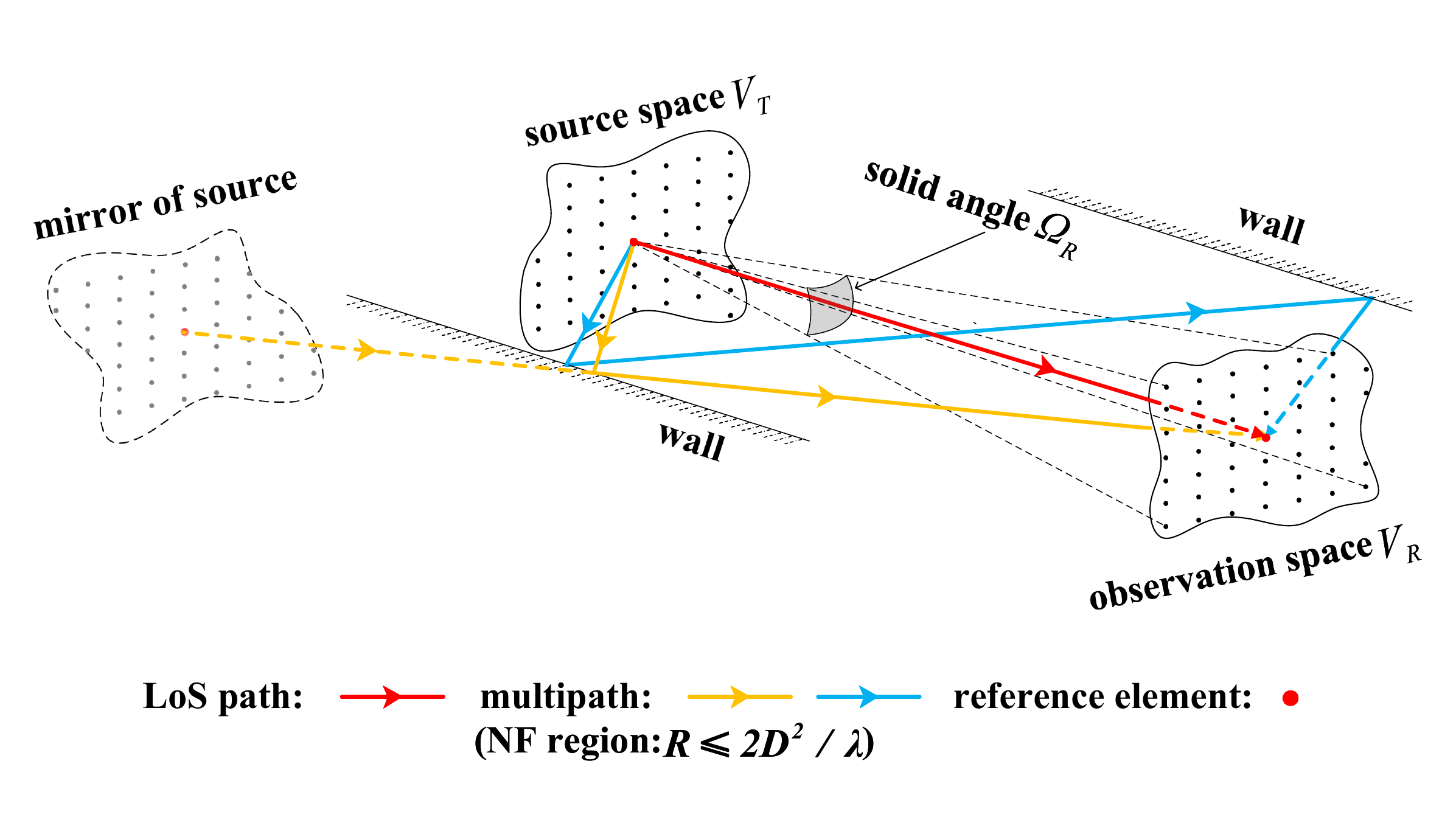}
\caption{NF multipath propagation channel in UM-MIMO deployment. The mirror of the source denotes a virtual source for the specularly reflected path (i.e., the yellow line) according to the image-propagation mechanism \cite{bertoni1999radio}, and the shaded area represents the solid angle illuminated by the projection of the Rx array onto the Tx array along the LoS path.}
\label{system_model}
\end{figure}

\section{System Model and Problem Statement}
\label{sec:system_model}
Consider a wideband UM-MIMO deployment as shown in Fig.~\ref{system_model}, where a transmitter (Tx) equipped with $N_T$ antennas (distributed in the source space $V_T$) communicates with a receiver (Rx) containing $N_R$ antennas (in the observation space $V_R$). Without loss of generality, we assume that the Tx and Rx arrays are exactly opposite each other along the LoS direction, and that the antennas are isotropic. The position vectors of the $n$th Tx element and the $m$th Rx element are $\mathbf{r}_{T,n} \in V_T$ and $\mathbf{r}_{R,m} \in V_R$, respectively. 
For such large Tx and Rx arrays, NF propagation occurs when the Tx-Rx distance $R$ is less than the FF Rayleigh distance defined as $2D^2/\lambda$, with $D$ denoting the array aperture and $\lambda$ representing the wavelength. Specifically, the waves propagate with a spherical wavefront in the NF propagation condition \cite{yuan2023phase}. Besides, as introduced before, practical UM-MIMO channels exhibit the multipath propagation characteristic, with other paths (such as reflections) existing in addition to the LoS path due to the interaction of the wave with objects around Tx and Rx.

Considering a specific subcarrier with the frequency $f$, the signal model is formulated as 
\begin{equation}
    \mathbf{y}(f) = \mathbf{G}(f)\mathbf{x}(f) + \mathbf{n},
    \label{eq:mimo_io}
\end{equation}
where $\mathbf{x}(f) \in \mathbb{C}^{N_T}$ and $\mathbf{y}(f) \in \mathbb{C}^{N_R }$ represent the transmitted and received signal vectors, respectively, $\mathbf{n}\in \mathbb{C}^{N_R }$ denotes the additive Gaussian white noise (AWGN) vector. Moreover, $\mathbf{G}(f) \in \mathbb{C}^{N_R \times N_T}$ is the channel transfer matrix, representing channel responses that determine the input-output relationship. In the following, we investigate channel modeling first for only the LoS transmission and then for multipaths in practical scenarios, further characterizing the transfer matrix.


\subsection{NF Channel Modeling}
\label{subsec:signal_model}
\indent\textit{1) For LoS transmission:}
With a monochromatic source distribution $\psi(\mathbf{r})$ within the space $V_T$ that generates waves $\phi(\mathbf{r})$, the channel transfer function $\mathbf{G}$ is physically constrained by the inhomogeneous Helmholtz equation \cite{miller2000communicating}
\begin{equation}
\nabla^2 \phi(\mathbf r) + k^2 \phi(\mathbf r) = -\psi(\mathbf r), 
\label{eq:helmholtz}
\end{equation}
with 
\begin{equation}
\phi(\mathbf{r}_R) = \int_{V_T} G(\mathbf{r}_R, \mathbf{r}_T) \psi(\mathbf{r}_T) \mathrm{d}\mathbf{r}_T.
\label{eq:integral_sol}
\end{equation}
The Green's function for eqn. (\ref{eq:helmholtz}), i.e., the solution that decides the waves at position $\mathbf{r}_{R,m}$ resulting from a point source at position $\mathbf{r}_{T,n}$ in the LoS propagation, is known as \cite{miller2000communicating} 
\begin{equation}
g_{m, n}(f)= \frac{e^{-j 2\pi f \|\mathbf{r}_{R,m} - \mathbf{r}_{T,n}\|/c}}{4\pi \|\mathbf{r}_{R,m} - \mathbf{r}_{T,n}\|},
\label{eq:greens_func}
\end{equation}
where $g_{m,n}(f)$ denotes the $(m,n)$th entry of $\mathbf{G}(f)$, and $c$ is the speed of light. 

\indent\textit{2) For multipaths:}
As mentioned before, the channel in practical UM-MIMO deployment scenarios exhibits the NF multipath propagation property. In such a case, the channel response $g_{m,n}(f)$ is the superposition of the LoS path and all other paths, as
\begin{equation} 
g_{m, n}(f) = g_{m, n}^{\mathrm{LoS}} (f)+ \sum_{l=2}^{L}g_{m, n}^{l}(f).
\label{eq:h_total} 
\end{equation} 
where $L$ is the number of paths in the channel. From the perspective of parametric channel modeling, $g_{m, n}(f)$ can be re-written leveraging the channel multipath parameters, as \cite{molisch2006comprehensive},
\begin{equation}
    {g}_{m, n}(f) = \sum_{l=1}^{L} \alpha_l e^{-j2\pi f \tau_l} s_{m,n}(f;\mathbf{\Theta}_{l}),
    \label{ISP Gen}
\end{equation}
with 
\begin{equation}
     s_{m,n}(f;{\mathbf{\Theta}}_{l})= \frac{d_l}{d_{l,m, n} } e^{-j2\pi f(d_{l,m, n} - d_l)/c},
    \label{ISP Gen1}
\end{equation}
where $s_{m,n}(f;\mathbf{\Theta}_l)$ represents the element of the array steering vector. The channel multipath parameters are defined with reference to a specific Tx-Rx array element pair (as shown in Fig. \ref{system_model}). The parameters $\{\alpha_l,\tau_l\}$ denote the complex amplitude (i.e., capturing the path amplitude attenuation during free-space propagation and bounces with scatterers) and the propagation delay of the $l$th path, respectively, and $\mathbf{\Theta}_l = \{ \theta_{l,\text{AoA}}, \theta_{l,\text{AoD}}, d_l\}$ denotes the spatial parameters. Specifically, $\theta_{l,\text{AoA}/\text{AoD}}$ is the angle of arrival/departure of the path to the Rx/Tx side, and $d_l$ denotes the sum of two distances at both ends of the spherical propagation, i.e., the one between the Tx reference element and the first-lap scattering source and the other one between the last-lap scattering source to the Rx reference element \cite{yin2017scatterer}. Moreover, $d_{l,m,n}$ denotes the sum of the two distances between the $(m,n)$th elements and the scattering sources, which can be expressed by $\mathbf{\Theta}_l$ considering array geometries and path propagation angles. Note that (i) the LoS transmission, i.e. (\ref{eq:greens_func}), is a special case of (\ref{ISP Gen}), with $\{\alpha_l, \tau_l, d_l,\theta_{l,\text{AoA}/\text{AoD}}\}_{l=1} = \{1/(4\pi R), R/c, R, 0^\circ\}$, and (ii) for LoS and reflected paths, $d_l = c\tau_l$ holds \cite{yin2017scatterer}.

\subsection{Existing Metric for NF DoF Calculation}
\label{subsec:metrics}
Although spatial DoF in FF cases has been thoroughly investigated (e.g., \cite{poon2005degrees}), most existing studies on NF DoF focus on LoS transmission (i.e., channels with only the LoS path) \cite{miller2000communicating,pizzo2022landau,do2022dof,kanatas2024degrees,chen2025effective}.
These studies investigate the eigenvalue distributions of the channel spatial correlation $\mathbf{K} = \mathbf{G}^\mathrm{H}\mathbf{G}$ and then extract the DoF metric. Specifically, by performing eigenvalue decomposition (EVD) on $\mathbf{K}$ with (\ref{eq:greens_func}) (i.e., the NF LoS transmission), the eigenvalues are derived under the constraint of the defining eigenfunction for the prolate spheroidal wave function \cite{frieden1971viii}. With a prior knowledge of the function property, the eigenvalue distribution can be obtained, and a critical value after which the eigenvalues fall off rapidly is extracted and referred as the DoF \cite{miller2000communicating}. Finally, the spatial DoF of an NF channel with only the LoS path is derived as the product of the electrical aperture projection and the channel solid angle, mathematically (with arrays perpendicular to the LoS path), 
\begin{equation}
n_\mathrm{DoF} = \frac{D_R \Omega_T}{\lambda^\eta} = \frac{D_T \Omega_R}{\lambda^\eta} = \frac{D_R D_T}{(\lambda R)^\eta},
\label{core：formula}
\end{equation}
where $\lambda$ is the wavelength, $D_T$ and $D_R$ denote the aperture projections of Tx and Rx arrays, respectively, $\Omega_T=D_T/R^\eta$ and $\Omega_R = D_R/R^\eta$ represent the solid angles, and $\eta$ denotes the dimensionality of arrays. An interpretation of (\ref{core：formula}) is that the DoF corresponds to the number of spatially resolvable beams, obtained by dividing the channel solid angle $\Omega_T = D_T / R^\eta$ by the spatial resolution $1/(D_R / \lambda^\eta)$ (related to the electrical array aperture). In practical implementations for uniform linear, circular, and planar arrays (ULA, UCA, and UPA), $\{D,\eta\}$ are equal to \{the length, 1\}, \{the diameter, 1\}, and \{the array area, 2\}, respectively.

\begin{figure}[t!]
\centering
\includegraphics[width=3.2in]{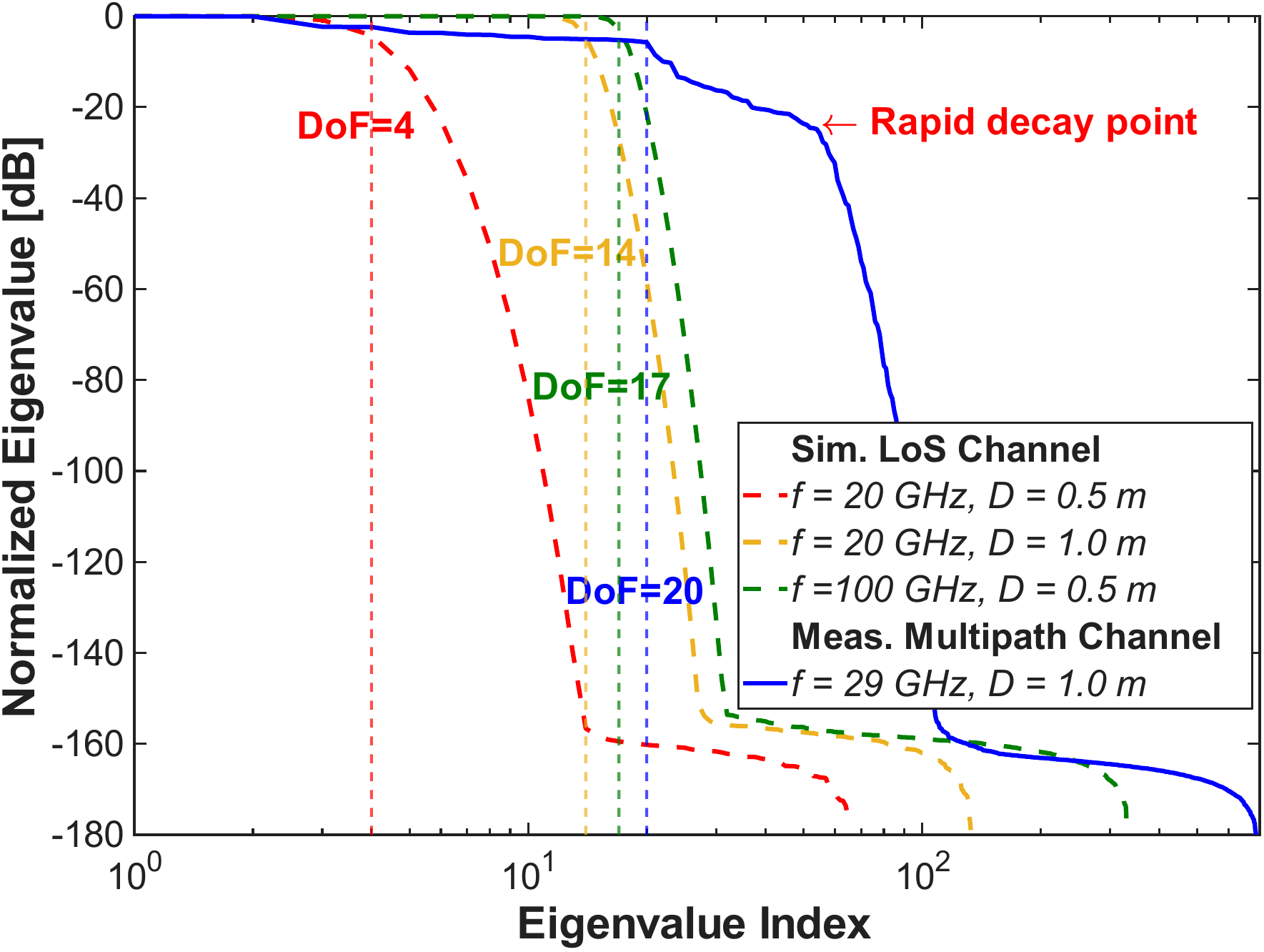}
\caption{
Normalized eigenvalue distributions of various simulated ULA-ULA channels with LoS path only and a measured ULA-UCA multipath channel. The Tx-Rx distance is fixed with $R$ = 5 m. The measured channel, with measurement configurations found in the later section \ref{subsec:mea_cam}, denotes a LoS multipath channel with 9 paths (shown in Fig. \ref{final results}). The marked DoF values are calculated according to the existing metric, i.e., (\ref{core：formula}).}
\label{formula}
\end{figure}

\subsection{Problem Statement}
While the existing metric for NF DoF only considers the LoS transmission, multipaths exist in practical UM-MIMO deployed channels. Therefore, questions of whether and how much the DoF is affected (increased or decreased) arise naturally, which remains a gap in the literature. Fig.~\ref{formula} illustrates eigenvalue distributions of simulated channels with only a LoS path, and that of a measured NF multipath channel, obtained via EVD of the spatial correlation matrix $\mathbf{K} $. It can be observed that all eigenvalue curves initially remain unchanged or slightly decrease and then drop sharply, where the rapid decay points are known as the DoF. As shown in Fig. \ref{formula}, the existing DoF metric, i.e., (\ref{core：formula}), can accurately calculate the DoF for NF LoS-path channels, and the DoF is increased relative to the FF LoS-path channel (where DoF equals 1). When considering the measured multipath channel, however, the existing metric fails to calculate the channel DoF. A significant discrepancy between the calculated DoF and the empirical one is observed, leading to an underestimation. This result underscores the limitations of the existing metric in capturing the DoF of NF multipath channels. Consequently, a generic DoF metric for practical NF multipath scenarios is required.

\section{Proposed DoF Metric for NF Multipath Channels}
\label{third}
Apart from a singular LoS path, an NF UM-MIMO channel typically contains multipaths resulting from interactions with reflective surfaces such as walls and floors. This characteristic is particularly critical in NLoS scenarios, where reflected paths provide the primary and often the only means of connectivity.
In this section, we first propose a novel and effective DoF metric for NF multipath channels. Reflected paths, as dominant multipath components, are modeled based on the concept of a virtual Tx (VTx) and incorporated into the DoF analysis through theoretical derivation. Then, how to implement the proposed metric is detailed, followed by a discussion of its applicable scenarios and limitations.

\subsection{Generic DoF metric}
\label{subsec:metric}

We consider a multipath scenario consisting of a LoS path and multiple reflected paths (including multi-order reflections). Note that in this analysis, other propagation mechanisms, i.e., diffraction, penetration, and diffuse, are ignored, as they are usually with relatively weaker power and hence not dominant in UM-MIMO systems \cite{yuan2022spatial}. According to the image-propagation mechanism \cite{bertoni1999radio}, paths undergoing LoS transmission or reflections will correspond to spherical waves whose center is the Tx or Tx mirror (i.e., VTx), as shown in Fig.~\ref{system_model}. Hence, with $L$ paths, it is actually equivalent to the Rx receiving waves from $L$ spatially distributed VTx. In such a case, reflected paths undergo only a longer propagation distance (affecting the delay) and an additional attenuation (known as the reflection coefficient). Hence, for the $l$th reflected path, the channel response $g_{m, n}^{l}(f)$ can be re-written as
\begin{equation}
    g_{m, n}^{l }(f)= \Gamma_l \frac{e^{-j 2\pi f \|\mathbf{r}_{R,m} - \mathbf{r}_{T,n}^\prime\|/c}}{4\pi \|\mathbf{r}_{R,m} - \mathbf{r}_{T,n}^\prime\|}
    \label{eqn:ref_for}
\end{equation} 
where $\Gamma_l $ denotes the reflection coefficient during the reflections \cite{degli2011analysis} and $\mathbf{r}_{T,n}^\prime$ denotes the position of the $n$th element in VTx. 

With such modeling for the reflected path, its DoF can be intuitively regarded as that of a LoS path between the Rx array and the VTx array. Similar to the derivation for the LoS path in \cite{poon2005degrees}, the DoF of the reflected path can be derived as the product of the effective aperture projection and the channel solid angle (corresponding to VTx and Rx), mathematically,
\begin{equation}
    n_\mathrm{DoF}^l = \frac{D_l^{\text{eff}} \Omega_l^{\text{eff}}}{\lambda^\eta} = \frac{(D_T \cos \theta_{l,\text{AoD}})  (D_R \cos \theta_{l,\text{AoA}})}{(\lambda d_l)^\eta} \quad 
    \label{eqn:singleDoF}
\end{equation}
where $D_l^{\text{eff}}$ and $\Omega_l^{\text{eff}}$ signify the effective aperture area and the subtended solid angle for the $l$th path, respectively. 

Let us conclude the channel DoF with all paths considered. A straightforward approach is to directly summarize the DoF contributions from each individual path as the channel DoF. However, some paths are projected with an overlap in the solid angle, due to the spatial resolution limitation of the arrays. For example, paths that share the same azimuth angle but have different elevation angles are indistinguishable from a horizontal ULA, hence resulting in an overlap in the projected solid angle. Consequently, the DoF of the multipath channel converges to the effective union of these spatial DoF from each path, as
\begin{equation}
    N_\mathrm{DoF}^\mathrm{Total} = \left\lfloor \bigcup_{l=1}^{L} n_\mathrm{DoF}^l\right\rfloor,
    \label{eqn:doftotal}
\end{equation}
where the operation $\bigcup$ represents effective summation, which directly eliminates the DoF contribution caused by solid angle overlap to avoid repeated superimposition, and the operation $\lfloor \cdot \rfloor$ aims at an integer of the DoF. Note that the overlap can be determined via the array geometry and estimated path parameters in practical implementation (see Section \ref{subsec:met_impl}). This metric inherently accounts for the DoF gains harvested from spatially resolvable multipaths while simultaneously capturing the saturation effects induced by spatial overlaps. 

We further reveal the mapping relationship between the eigenvalues associated with the path DoF and the corresponding path powers. In FF cases, by setting the eigenvectors in the form of steering vectors (which are orthogonal in the FF case, e.g., discrete Fourier transform), the channel eigenvalues from the EVD can be derived to be equal to the powers of the discriminable paths. This explains why the number of resolvable multipaths coincides with the spatial DoF in FF channels when sufficiently large arrays are employed. In NF cases, however, a single path may span multiple spatial modes, as mentioned above. By deriving the eigenvalue distribution via the prolate spheroidal wave function (see Section \ref{subsec:metric} for details) based on the green function operators for each path, i.e., (\ref{eq:greens_func}) and (\ref{eqn:ref_for}), it can be easily obtained that the sum of eigenvalues within the DoF region for each (non-overlapping) path equals the corresponding path power, i.e.,
\begin{equation}
    |\alpha_l|^2 = \sum_{i\in \Omega_l}^{} \lambda_i,
    \label{eqn:relat}
\end{equation}
where $\Omega_l$ denotes the path DoF range. Note that the used eigenvalues here are normalized by the array gains, i.e., $10 \lg(N_T N_R)$. This mapping further provides practical guidance for the metric implementation under power-constrained conditions (e.g., low signal-to-noise ratio (SNR)), indicating which DoF should be activated according to the water-filling principle \cite{tse2005fundamentals}.

\subsection{Metric implementation}
\label{subsec:met_impl}
\textbf{Generic implementation}: The proposed metric can be implemented using the multipath propagation parameters $\{\alpha_l, \tau_l, \theta_{l,\text{AoA}}, \theta_{l,\text{AoD}}\}$ and system configurations $\{D,\eta,\lambda\}$, which can be estimated or are typically available in UM-MIMO deployments. First, the DoF contributed by each path is calculated according to (\ref{eqn:singleDoF}) based on the array dimensions $\{D,\eta\}$, wavelength $\lambda$, and path parameters $\{\tau_l,\theta_{l,\text{AoA}},\theta_{l,\text{AoD}}\}$ (with $d_l=c\tau_l$ for LoS and reflected paths). By accounting for spatial overlaps determined by the array geometry and propagation trajectories, the individual path DoF are then aggregated to obtain the overall channel DoF via (\ref{eqn:doftotal}). This value represents the maximal number of sub-channels that can be exploited for data transmission in high-SNR regimes. In low-SNR regimes, however, the effective DoF may be smaller. Specifically, the eigenvalues associated with each path DoF, representing the sub-channel powers, can be determined from the above mapping based on the path amplitude $\alpha_l$. The effective DoF are then given by the eigenmodes whose powers exceed the water-filling threshold under the optimal power allocation.

\textbf{Applicable scenarios and limitation analysis}: The proposed metric is applicable to NF LoS-transmission channels, NF LoS multipath channels, and NF NLoS multipath channels, covering typical NF UM-MIMO deployment scenarios. For LoS-transmission channels, the existing NF DoF metric (\ref{core：formula}), which is demonstrated to be effective, can be directly obtained as a special case of the proposed metric (\ref{eqn:doftotal}) by setting the number of paths to $L=1$. For NF multipath channels in both LoS and NLoS scenarios, the metric remains applicable since dominant reflected paths are modeled through (\ref{eqn:ref_for}), and the corresponding path-wise DoF is derived accordingly. The validity of the proposed metric is further demonstrated through numerical simulations and experimental measurements in the following sections.
Regarding the metric limitation, only LoS transmission and specular reflections are assumed and considered in the analysis, and other mechanisms such as diffraction are neglected. Nevertheless, in UM-MIMO systems operating at high frequencies (e.g., mmWave and THz bands), LoS and specular reflections typically constitute the dominant propagation components for high-capacity transmission. Under such conditions, the proposed metric provides an effective approximation for practical DoF analysis.

\begin{figure}[t!]
\centering
\includegraphics[width=1\linewidth]{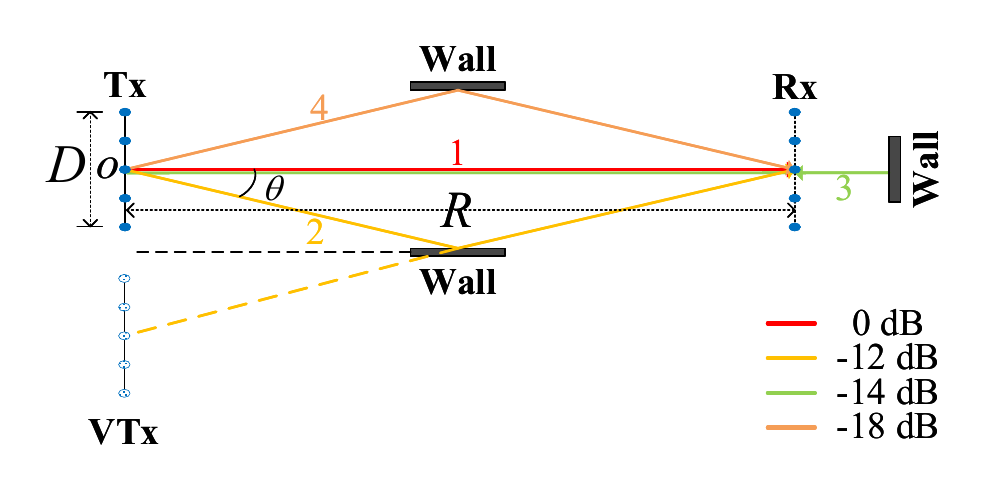}
\caption{Simulated NF multipath channel where 4 paths exist. The 3rd path, reflected by a wall, is incident on the Rx array at an angle opposite to the LoS path. The VTx corresponding to path 2 is drawn for demonstration.}
\label{simulated scenario}
\end{figure}

\section{Numerical Simulation} 
\label{fourth}
In this section, we perform a numerical simulation to demonstrate the proposed DoF metric for NF multipath channels. 
The simulation is configured in an NF UM-MIMO scenario at 29 GHz, as illustrated in Fig.~\ref{simulated scenario}. Both Tx and Rx are equipped with a ULA with an aperture of $D$ = 1 m and an element space of $\lambda/2$, and the distance between Tx and Rx is $R$ = 5 m. Note that the FF Rayleigh distance is calculated as 193.5 m, leading to NF propagation in the channel. The channel is simulated with a LoS path alongside three multipaths: two spatially resolvable specular reflections arriving at azimuth angles of $\theta = \pm 20^\circ$ with $d=5.32\,m$, and one back-reflected path $d=6\,m$. Note that from the side of Rx ULA, the back reflected path is completely overlapped and indistinguishable from the LoS path. Specifically, the power of the LoS path is normalized to 0 dB, while the powers of the two spatially resolvable reflections and the back-reflected path are set to -12, -18, and -14 dB, respectively. With those parameters, the channel is simulated according to (\ref{ISP Gen}) and (\ref{ISP Gen1}).

\begin{figure}[t!]
\centering
\includegraphics[width=0.97\linewidth]{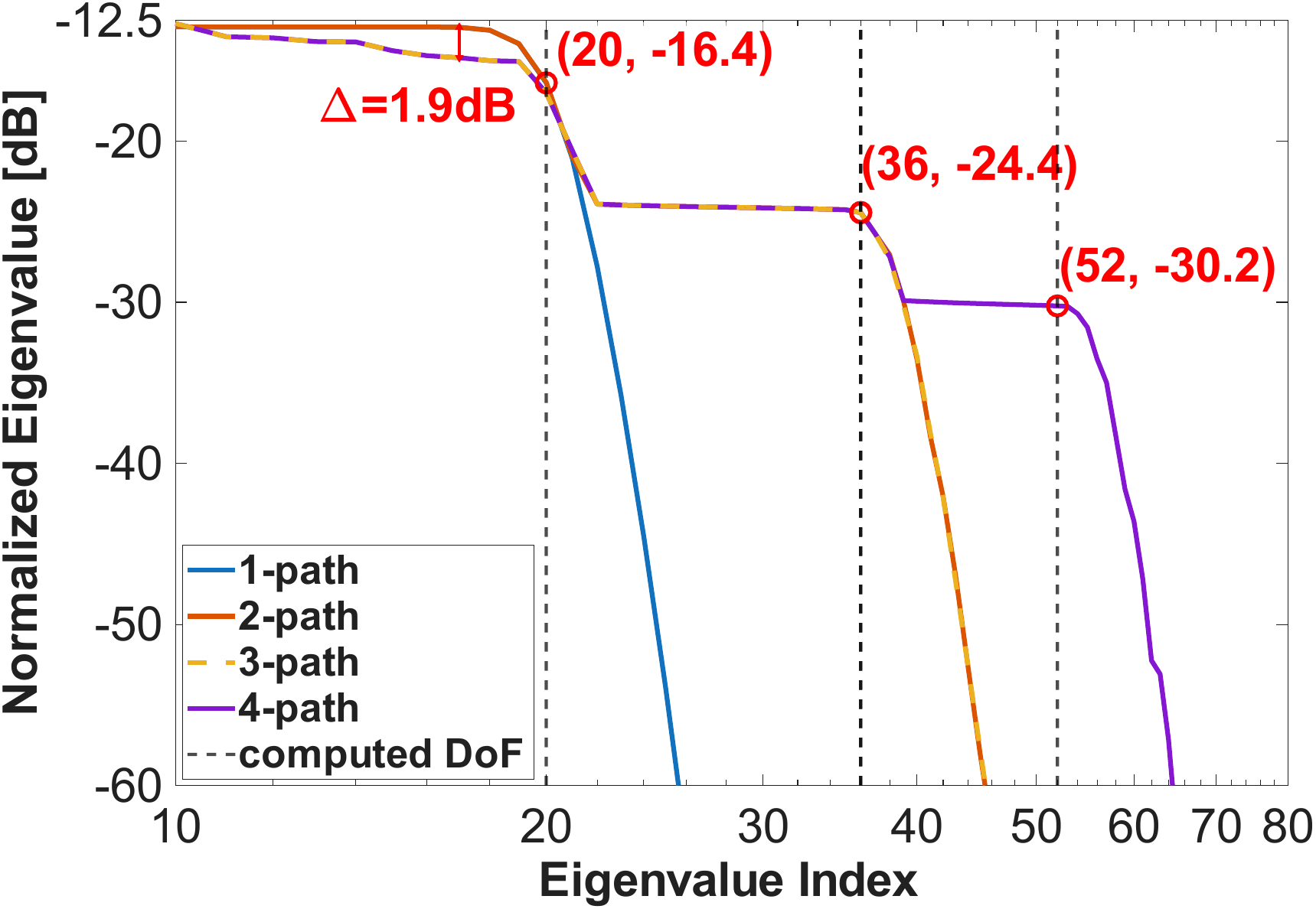}
\caption{
Eigenvalue distribution of the simulated NF multipath channel, normalized by the array gains. The legend of ``$l$th-path" denotes the multipath channel where the first $l$ paths ($1,2,....l$) are included.
}
\label{formula 1}
\end{figure}

Fig. \ref{formula 1} illustrates the eigenvalue distribution obtained via EVD performed on the simulated channels, and the DoF results are calculated according to the proposed metric, i.e., (\ref{eqn:doftotal}). Note that the illustrated eigenvalues are normalized by the array gains. Regarding both the LoS-transmission channel and multipath channels, i.e., cases of ``1-path" and ``2-path" or ``4-path", the calculated DoF aligns well with the target DoF extracted from the eigenvalue distribution (i.e., the rapid decay points). This result demonstrates the effectiveness of the proposed metric, not only for the NF LoS-transmission case but also for NF multipath channels. Note that the proposed metric degenerates into the existing metric (i.e., (\ref{core：formula})) for the ”1-path” case, indicating the existing metric is a special case of our proposed one. Besides, regarding the ``3-path" case, the eigenvalue distribution curve overlaps with that of the ``2-path" case, indicating no DoF increase from the back-reflected path. This occurs because the 3rd path and the LoS path are indistinguishable from the Rx ULA side and hence lead to the overlap of the solid angles. In other words, the two paths are regarded as one path from the Rx side. A 1.9 dB of eigenvalue deviation between cases of ``1-path" and ``3-path" is observed and marked in Fig. \ref{formula 1}. Such deviation is attributed to the superimposition of the LoS and 3rd paths (i.e., destructive summation) \cite{bohagen2007design}, indicating the influence of the overlapping. 

\begin{table}[t!]
\centering
\caption{Mapping relationship between eigenvalues and path power}
\renewcommand{\arraystretch}{1.2} 
\begin{tabularx}{\columnwidth}{c Y Y Y} 
\toprule
\textbf{Path $l$} & 
\textbf{Path power [dB]} & 
\textbf{Calculated DoF} &
\textbf{$\sum_{i\in \Omega_l} \lambda_i$ [dB]} \\ 
\midrule
1 & 0     & 20  & 0.1 \\ 
2 & -12 & 36 & -11.8 \\
3 & -14 & 36 & --- \\
4 & -18 & 52 & -17.4 \\
\bottomrule
\end{tabularx}

\vspace{4pt}
{\footnotesize
\raggedright
$\lambda_i$ denotes the eigenvalue within the path DoF range $\Omega_l$.
\par
}
\label{tab:simulation}
\end{table}

Table \ref{tab:simulation} illustrates the calculated DoF and the extracted summation of eigenvalues. As shown, the specific DoF values are 20, 36, and 52 for the three cases, respectively, indicating that NF propagation can significantly enhance the DoF gain compared to FF channels. Moreover, the extracted summation of eigenvalues corresponding to each path range is observed to match well with the preset path power, demonstrating the effectiveness of (\ref{eqn:relat}). This match reveals the relationship between sub-channel powers in multiplexing and multipath powers in the NF channel, which can be used for providing insights and guidance on the metric implementation in low-SNR scenarios for resource optimization purposes. Those results demonstrate the effectiveness of the proposed metric for DoF analysis of a practical NF multipath channel. 


\begin{figure}[!t]
    \centering
    \subfloat[]{\includegraphics[width=0.98\linewidth]{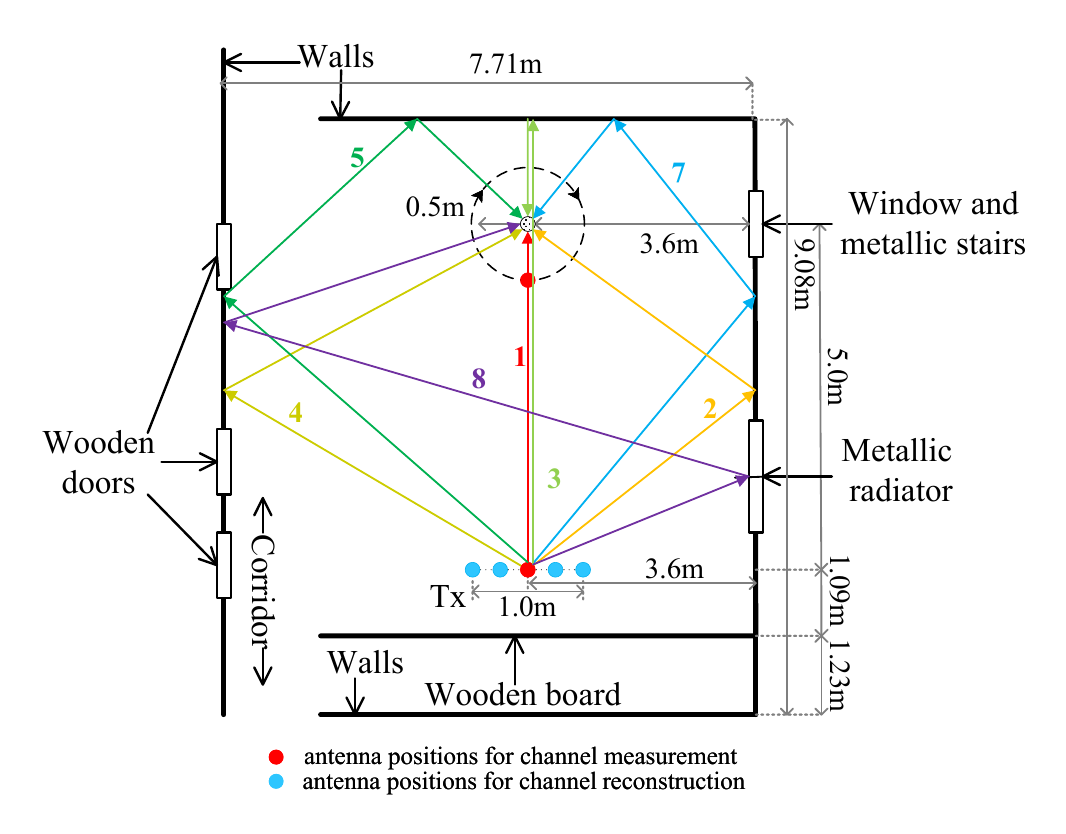}
    \label{subfig:LoS_meas_sce}}\\
    \subfloat[]{\includegraphics[width=0.98\linewidth]{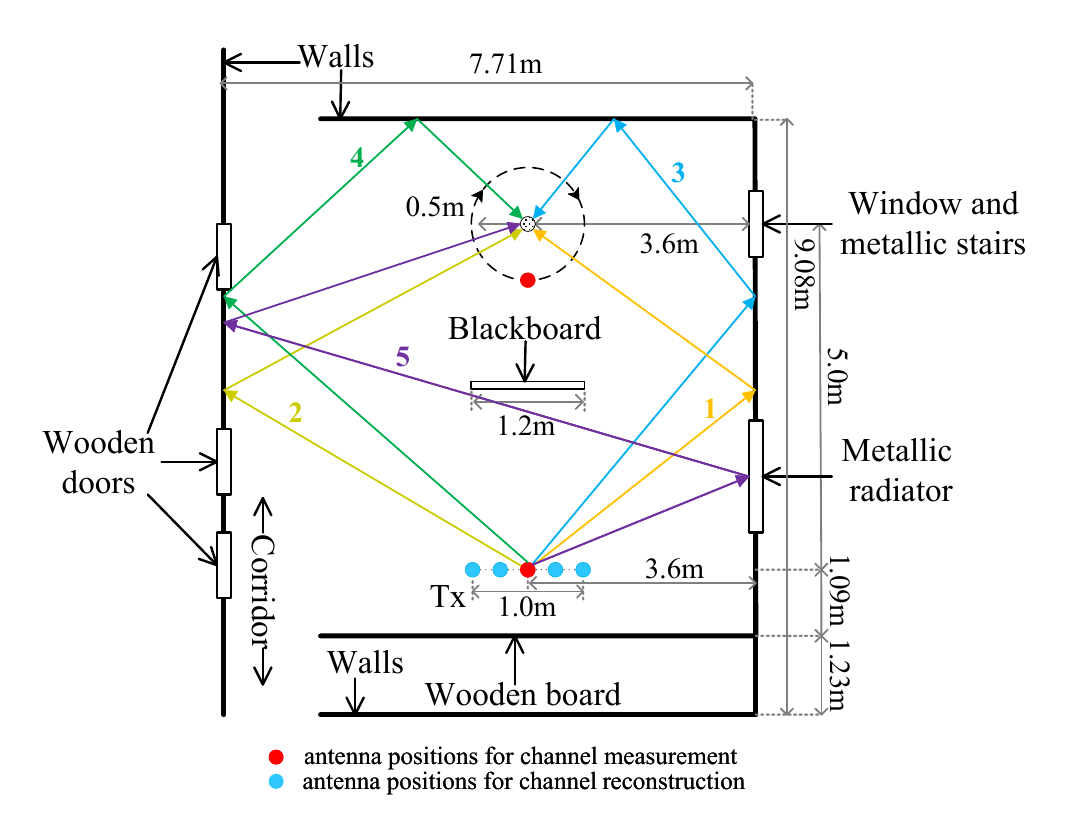}
    \label{subfig:NLoS_mea_sce}}\\
    \subfloat[]{\includegraphics[width=0.47\linewidth]{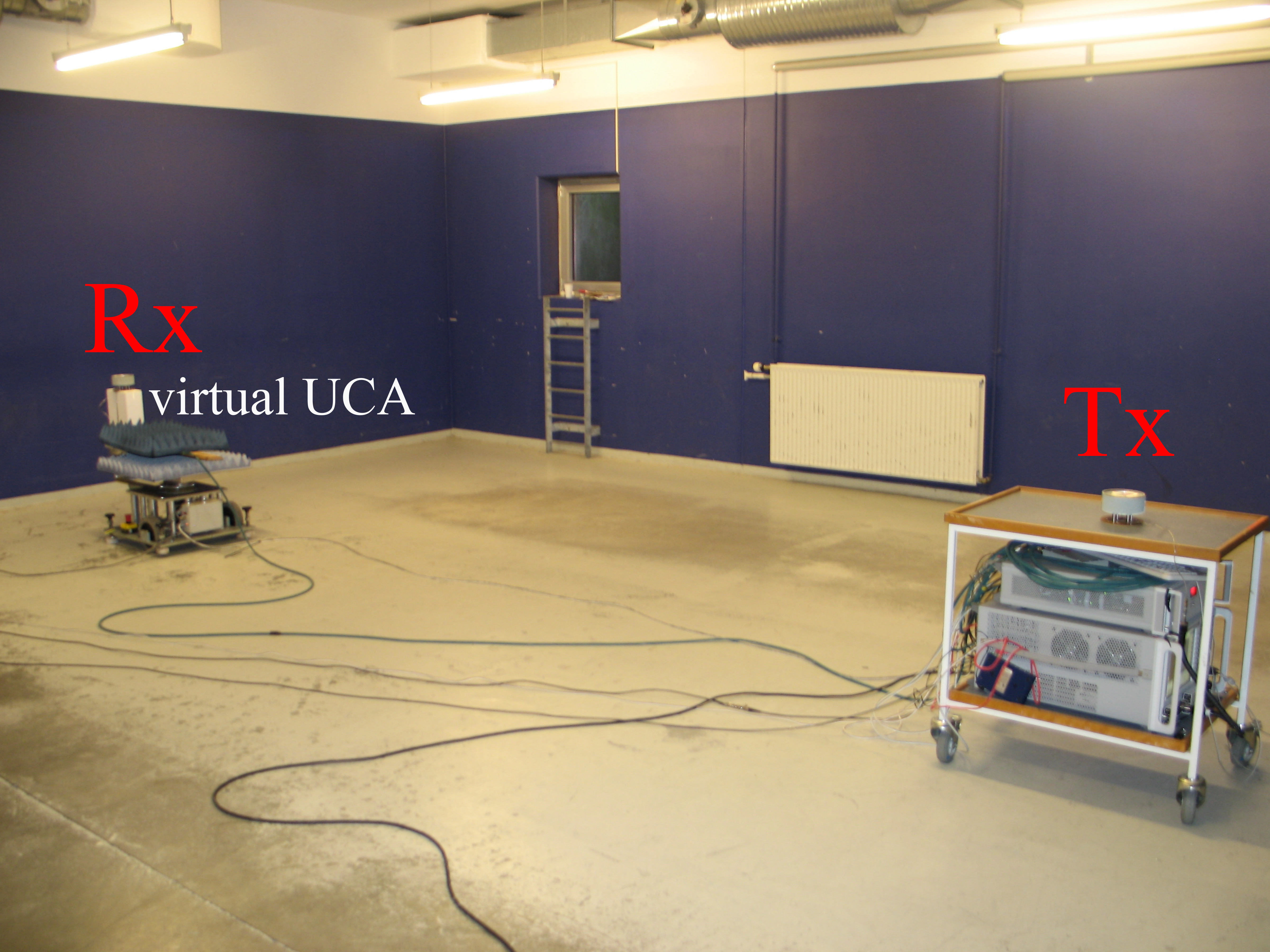}
    \label{subfig:LoS_pho}}
    \hfill
    \subfloat[]{\includegraphics[width=0.47\linewidth]{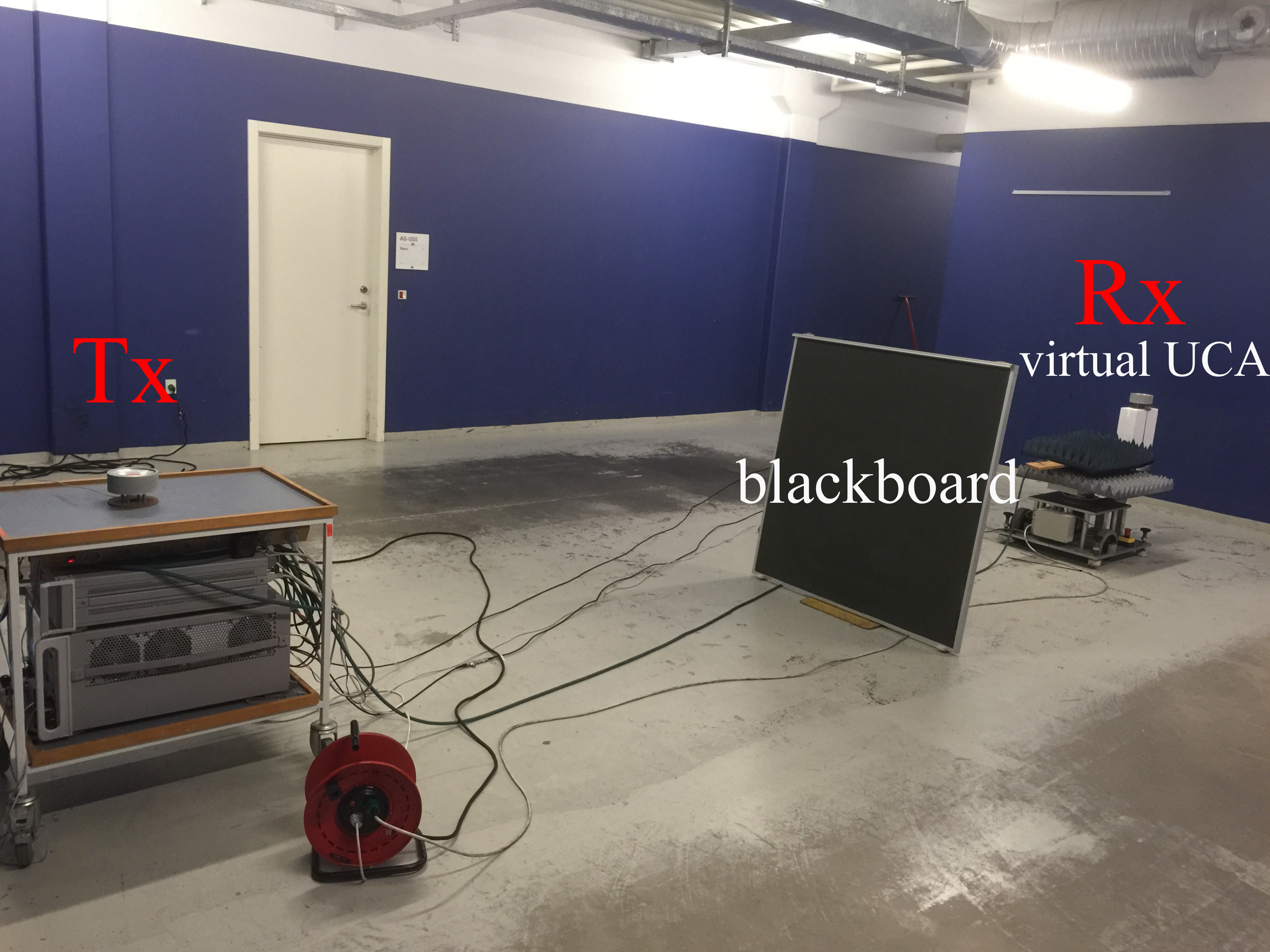}
    \label{subfig:NLoS_pho}}
    \caption{Illustrations of the indoor NF channel measurements, including (a) the LoS scenario, (b) the NLoS scenario (with a board between the Tx and Rx), (c) a photo for the LoS measurement, and (d) a photo for the NLoS measurement. The paths identified by the measurement results are marked in the scenarios. In (a), the 6th and 9th paths corresponding to ground and ceiling reflections, respectively, are not present due to the vision redundancy.}
    \label{measured scenario}
\end{figure}

\section{Experimental Validation}
\label{fifth}
In this section, we perform a real-world NF channel measurement and associated post-processing to validate the proposed NF DoF metric, including both the LoS and NLoS scenarios. First, we conduct an ultra-massive Single-Input Multiple-Output (UM-SIMO) measurement campaign using the virtual UCA scheme via a turntable (considering the low cost in both time and system hardware of the SIMO measurement and the virtual array scheme). Then, multipath parameters are estimated from the measured channels via a maximum likelihood estimation (MLE) algorithm \cite{ji2018channel}, and then are utilized to reconstruct the UM-MIMO channels. Finally, we extract the DoF values via the EVD on the reconstructed channels for various cases (single- and multiple-paths in LoS and NLoS scenarios), and compare them with the calculated DoF based on the proposed metric.

\begin{figure*}[!t] 
    \centering
    
    
    \subfloat[]{\includegraphics[width=0.33333\textwidth, trim=18 0 40 17,
      clip]{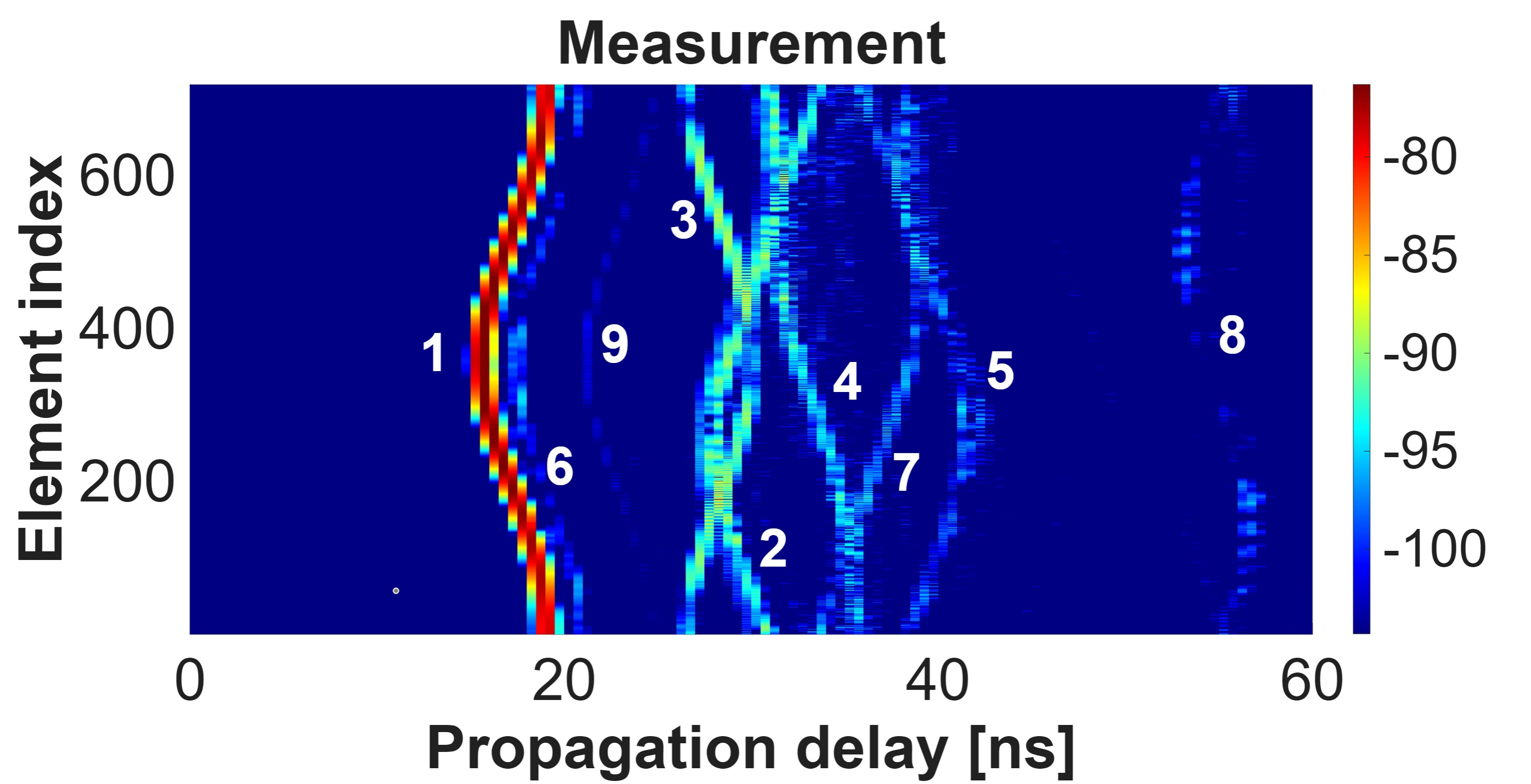}%
    \label{fig_first}}
    \hfil 
    \subfloat[]{\includegraphics[width=0.33333\textwidth, trim=18 0 40 17,
      clip]{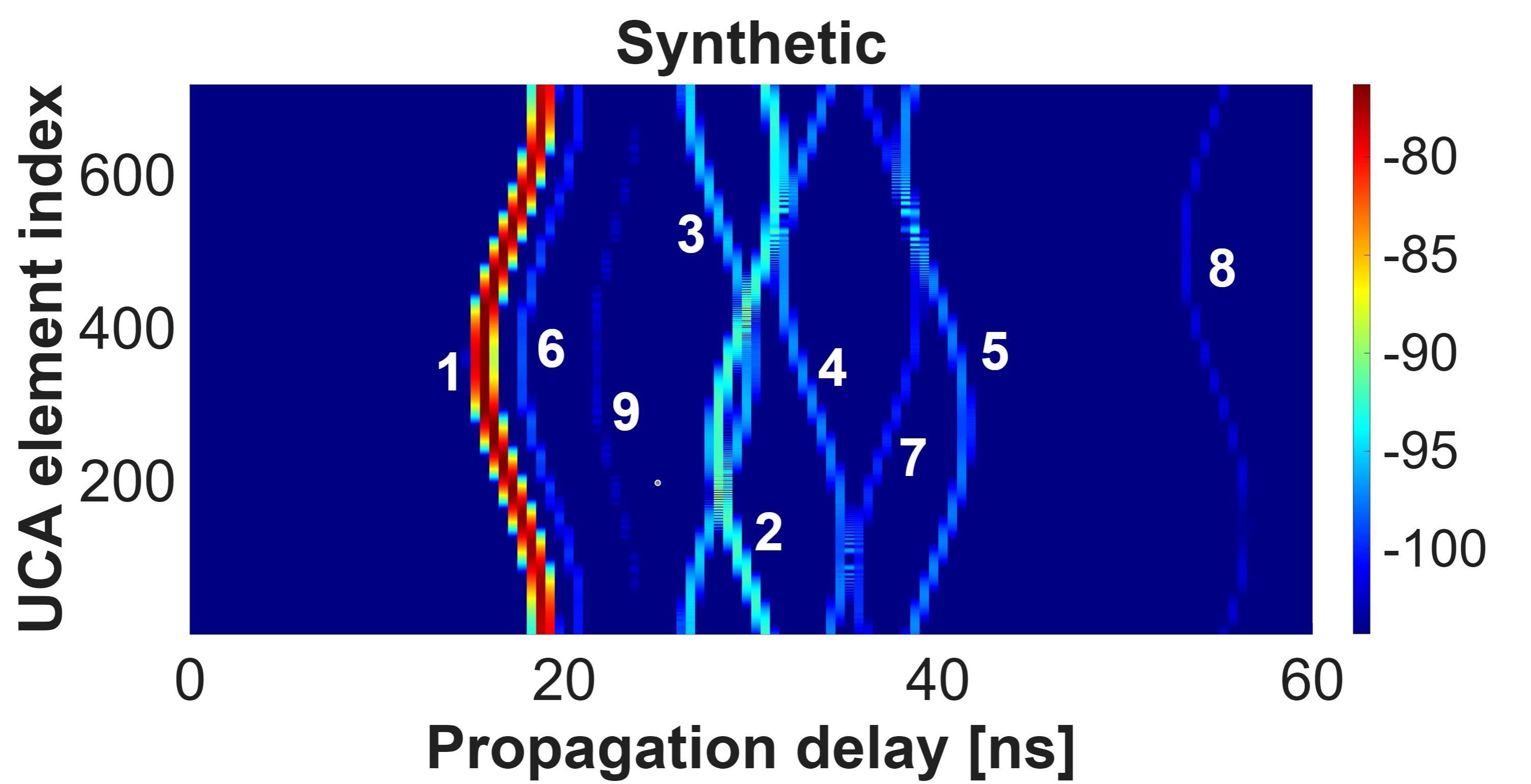}%
    \label{fig_second}}
    \hfil 
    \subfloat[]{\includegraphics[width=0.33333\textwidth, trim=18 0 40 17,
      clip]{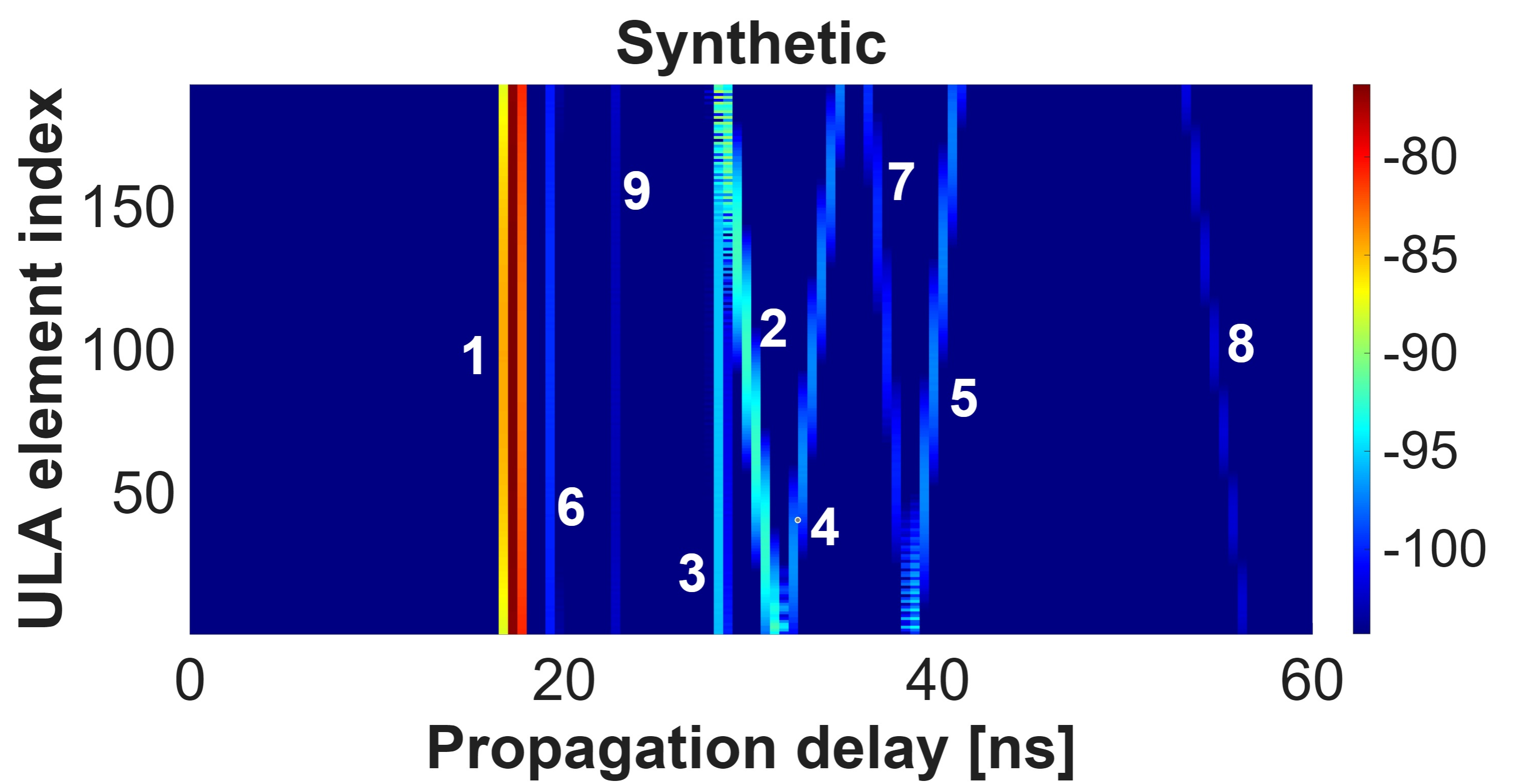}%
    \label{fig_third}}

    \caption{ Measured CIRs and synthetic CIRs based on extracted parameters in the LoS scenario.(a) Measured CIRs over virtual UCA elements.
(b) Synthetic CIRs over UCA elements (with the ULA center element as the Tx). 
(c) Synthetic CIRs over ULA elements(with the UCA center element as the Rx).}
    \label{fig_three_images}
\end{figure*}

\begin{figure*}[!t] 
    \centering
    
    
    \subfloat[]{\includegraphics[width=0.33333\textwidth, trim=18 0 40 17,
      clip]{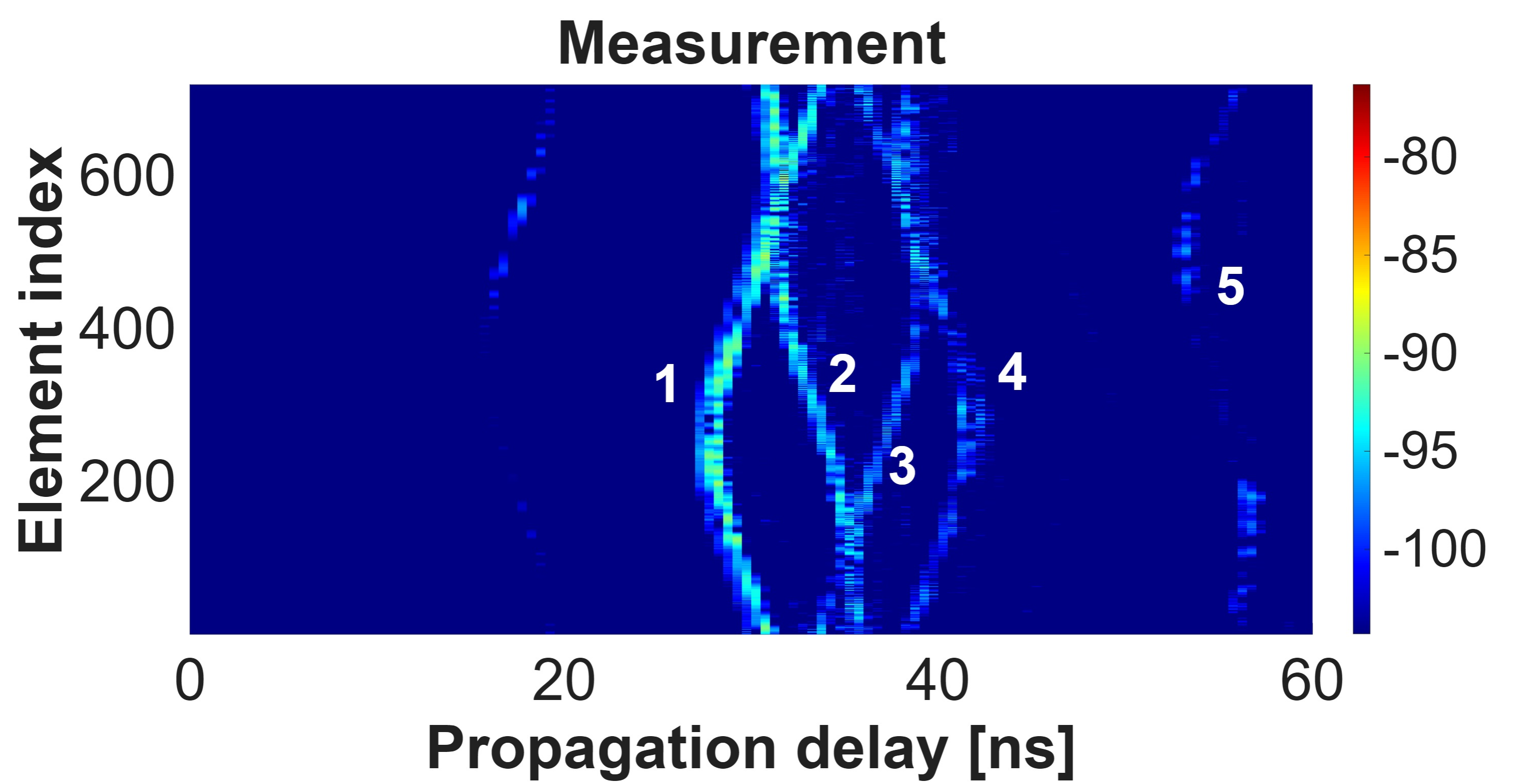}%
    \label{fig_first_nlos}}
    \hfil 
    \subfloat[]{\includegraphics[width=0.33333\textwidth, trim=18 0 40 17,
      clip]{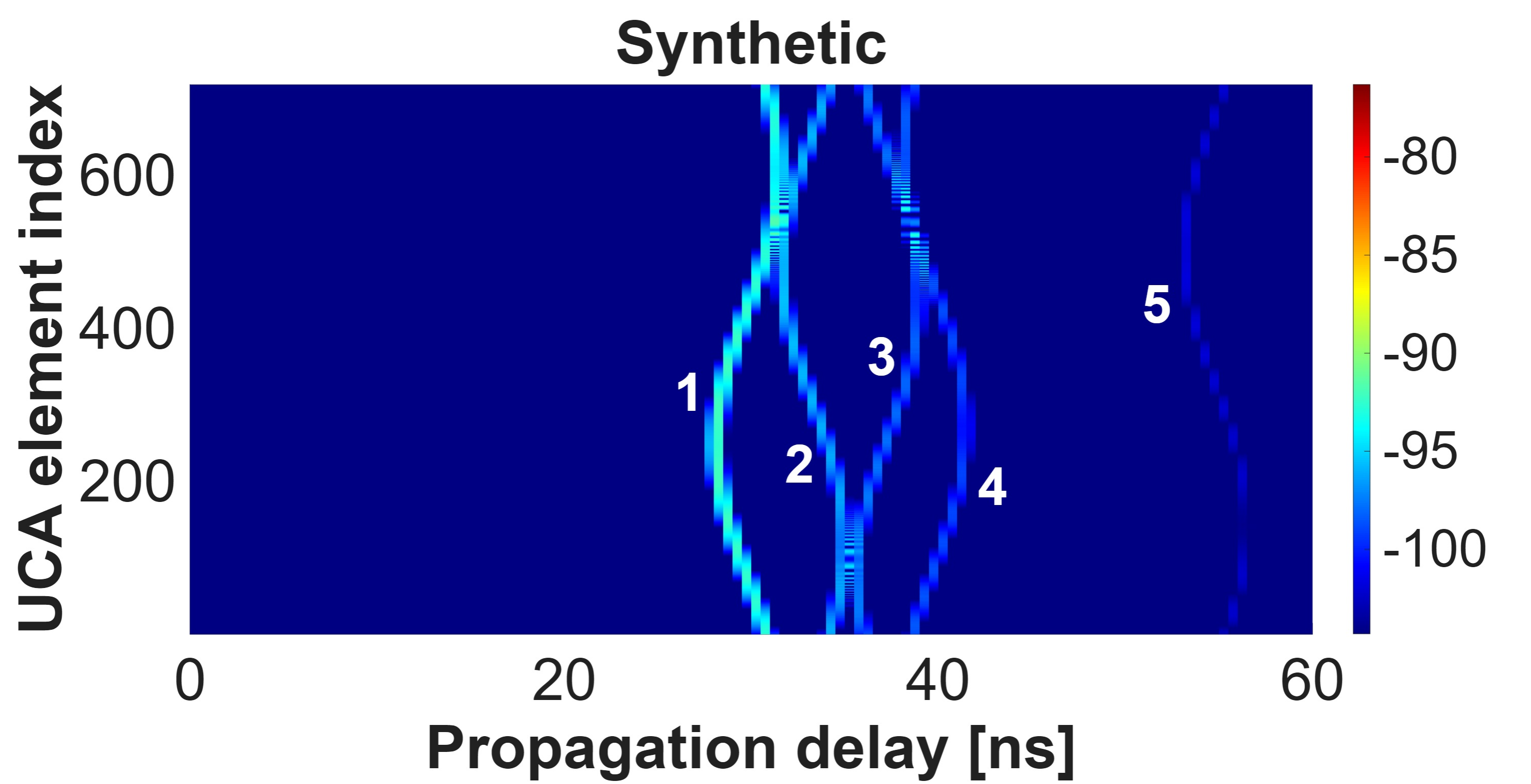}%
    \label{fig_second_nlos}}
    \hfil 
    \subfloat[]{\includegraphics[width=0.33333\textwidth, trim=18 0 40 17,
      clip]{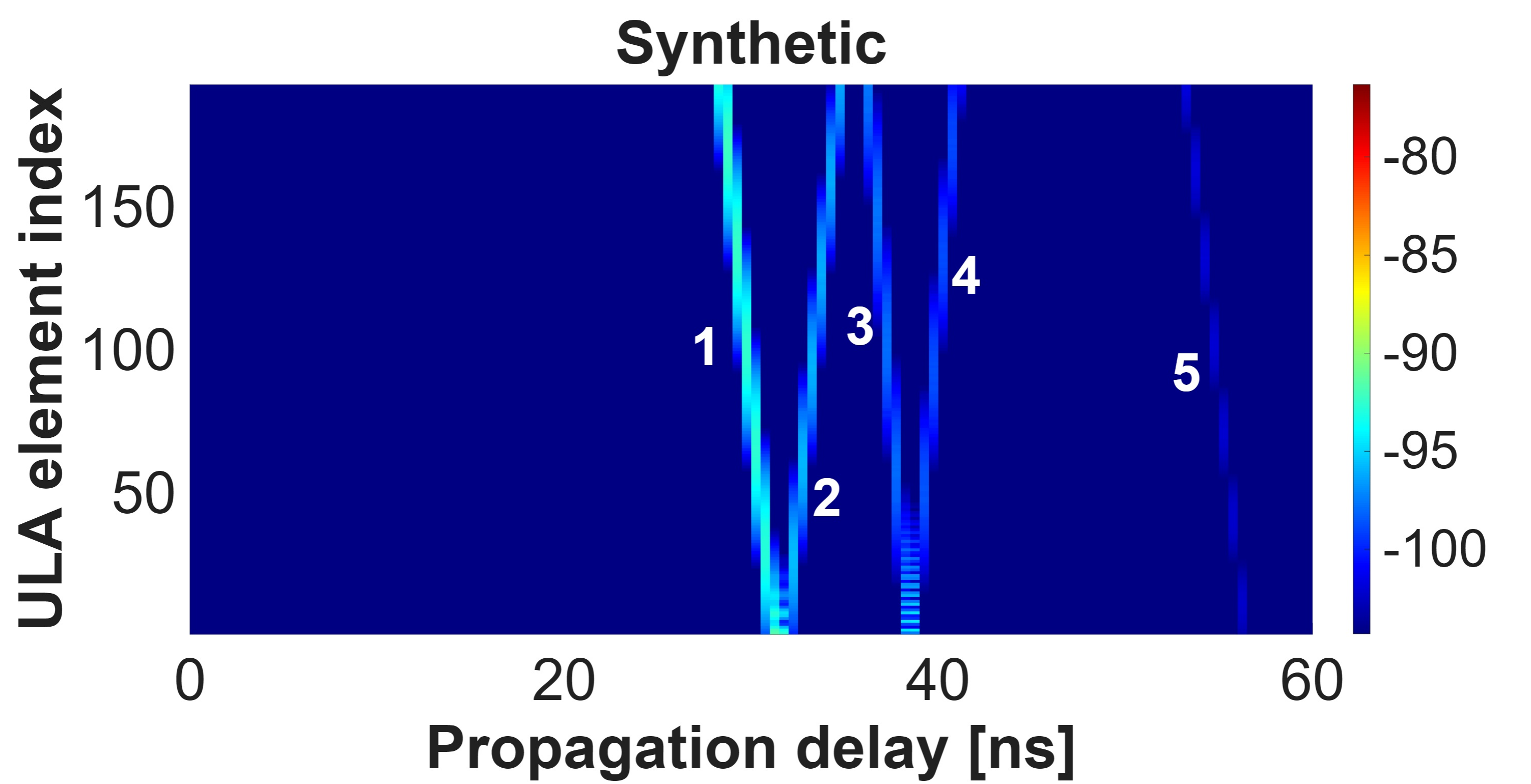}%
    \label{fig_third_nlos}}

    \caption{ Measured CIRs and synthetic CIRs based on extracted parameters in the NLoS scenario.(a) Measured CIRs over virtual UCA elements.
(b) Synthetic CIRs over UCA elements (with the ULA center element as the Tx). 
(c) Synthetic CIRs over ULA elements(with the UCA center element as the Rx).}
    \label{fig_three_images_nlos}
\end{figure*}

\subsection{Measurement Campaign}
\label{subsec:mea_cam}
An NF channel measurement campaign was conducted, including both LoS and NLoS scenarios, to validate the proposed metric, as illustrated in Fig.~\ref{measured scenario}. The measurements were carried out in an indoor room environment surrounded by objects such as walls, stairs, metallic heaters, and a whiteboard. The object materials and the room dimensions are identified in the figure. Wideband biconical antennas \cite{yuan2023phase}, featuring vertical polarization and a gain of 6 dBi in the 28–30 GHz band, were deployed in both Tx and Rx at a height of 0.84 m. The Tx-Rx distance was fixed at 5 m. A virtual UCA with a radius of 0.5 m was realized in Rx by rotating the antenna on a high-precision turntable, with a step of $0.5^\circ$ (in total 720 virtual elements). At each position, a vector network analyzer (VNA) was used to record the channel frequency responses, with 750 frequency points within 28-30 GHz. Note that the virtual array technique was adopted here due to its low cost and ease of realization. A blackboard was removed and put between the Tx and Rx arrays, resulting in the LoS and NLoS scenarios, respectively, as shown in Fig. \ref{subfig:LoS_meas_sce} and \ref{subfig:NLoS_mea_sce}. The power dynamic range of the measurement is set to 30 dB below to the LoS path power, aiming at a clear observation. Fig.~\ref{fig_first} and \ref{fig_first_nlos} present the measured channel impulse responses (CIRs, obtained via Discrete Fourier Transform of the measured channel frequency responses) over the virtual UCA elements. As shown, in addition to the LoS path, multipaths are observed in the channel. Those paths exhibit an ``s"-shaped curve in the element domain due to delay variations across elements. Within the same preset dynamic range of 30 dB, 9 paths and 5 paths are identified in the LoS and NLoS scenarios, respectively, and are marked in the figures.

\subsection{Parameter Extraction and Channel Reconstruction}
\label{parameter}
It is required to estimate multipath parameters from the measured UM-SIMO channel and then reconstruct the UM-MIMO channel for NF DoF analysis. Initially, a high-resolution MLE algorithm, which has been validated as implementable and effective for NF channel parameter estimation \cite{ji2018channel}, is adopted. Specifically, for the $l$th path, the parameters $\{\hat{\alpha}_{l}, \hat{\tau}_{l}, \hat{\mathbf{\Theta}}_{l}\}$ are obtained sequentially via an MLE matched-filtering process, and a successive interference cancelation process is performed to eliminate the $l$th path and then iteratively estimate the next path. The estimation for the $l$th path is formulated as \cite{ji2018channel}
\begin{equation}
\{ \hat{\tau}_{l}, \hat{\mathbf{\Theta}}_{l}\}=\arg \max_{{\tau}_{l}, {\mathbf{\Theta}}_{l}}\left\| z({\tau}_{l}, {\mathbf{\Theta}}_{l})\right\|,
\label{eq:parameter extraction}
\end{equation}
\begin{equation}
\hat{\alpha}_{l}=\frac{z(\hat{\tau}_{l}, \hat{\mathbf{\Theta}}_{l})}{\|\text{vec}\{\mathbf{a}(\mathbf{f};\hat{\tau_{l}},\hat{\mathbf{\Theta}}_{l})\}\|},
\label{eq:parameter extraction12}
\end{equation}
where
\begin{equation}
    z({\tau}_{l}, {\mathbf{\Theta}}_{l})=\frac{\text{vec}\{\mathbf{a}(\mathbf{f};\tau_{l},{\mathbf{\Theta}}_{l})\}^\text{H}\text{vec}\left\{\mathbf{Y}^l(\textbf{f}) \right\}}{\|\text{vec}\{\mathbf{a}(\mathbf{f};\tau_{l},{\mathbf{\Theta}}_{l})\}\|}
\end{equation}
with $\mathbf{Y}^l(\textbf{f}) = {\mathbf{G}}_{mea}  - \sum_{l'=1}^{l-1} {\mathbf{G}}(\textbf{f}; \hat{\alpha}_{l'}, \hat{\tau}_{l'},\hat{\mathbf{\Theta}}_{l'})$ denoting the measured channel responses after cancellation of the first $l-1$ paths. Note that here $\mathbf{f}$ denotes a frequency vector covering the sampling points in the bandwidth (assuming the number is $Q$). The item $\mathbf{a} \in \mathbb{C}^{N_R \times N_T \times Q}$ denotes the matching function, with its $(m,n,q)$th entry as $a_{m,n,q}(f;\tau_{l},{\mathbf{\Theta}}_{l})=e^{-j2 \pi f \tau_l}s_{m,n}(f;\mathbf{\Theta}_{l})$. It is also worth mentioning that the above processing can only estimate the incident angle on the Rx array side due to the SIMO measurements, while the exit angle on the Tx side is further estimated based on the extracted parameters. Specifically,  based on the ray optics, the specular reflection assumption, and the room geometry, we use the estimated incident angle to trace the path propagation and determine the trajectories by comparing the trajectory distance with the estimated delays. The extracted trajectories are marked in Fig. \ref{measured scenario}, showcasing consistency with the room geometry.

\begin{figure}[!t]
\centering
\includegraphics[width=3.2in]{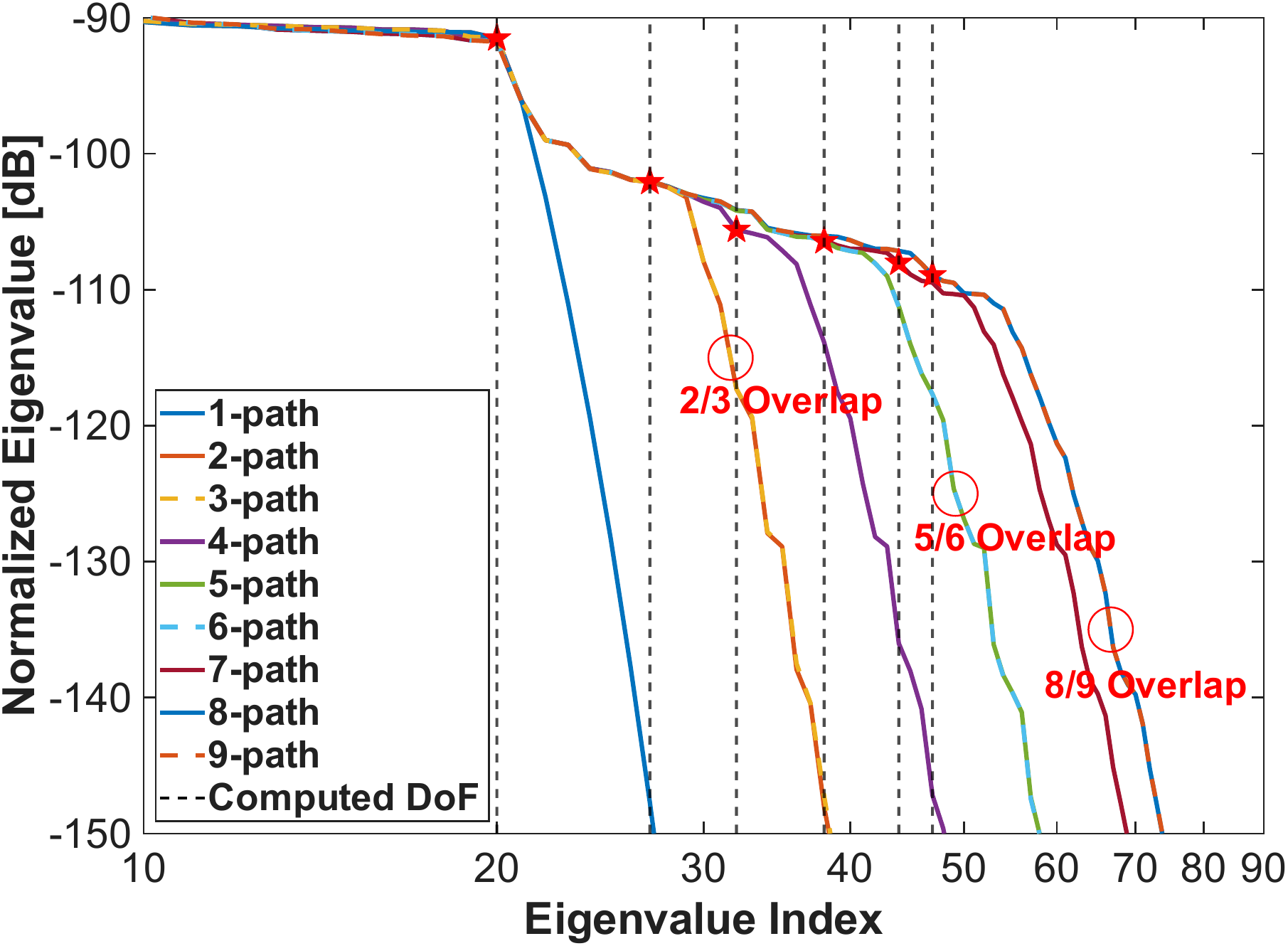}
\caption{Eigenvalue distribution of the synthetic ULA-UCA UM-MIMO channels, normalized by the array gains. The legend ``$l$th path" denotes the channel where the first $l$ paths ($1,2,....l$) exist. The red stars mark the interaction points between the computed DoF values and distribution curves (with specific values listed in Table~\ref{tab:path_dof}). Note that the dashed curves for the 3-,6- and 9-path overlap with the solid ones for the 2-,5- and 8-path, respectively, indicating no DoF increase for these paths due to the solid angle overlap.}
\label{final results}
\end{figure}

\begin{table}[!t]
\centering
\caption{Mapping relationship between eigenvalues and path power}
\renewcommand{\arraystretch}{1.2} 

\begin{tabularx}{\columnwidth}{c Y Y Y} 
\toprule
\textbf{Path $l$} & 
\textbf{Path power [dB]} & 
\textbf{Calculated DoF} &
\textbf{$\sum_{i\in \Omega_l} \lambda_i$ [dB]} \\ 
\midrule
1 & -76.4 & 20  & -76.4 \\ 
2 & -92.2 & 27 & -91.2 \\
3 & -95.3 & 27 & --- \\
4 & -97.1 & 32 & -96.2 \\
5 & -97.3 & 38  & -97.7 \\ 
6 & -99.6 & 38 & --- \\
7 & -99.9 & 44 &  -99.0\\
8 & -101.9 & 47 & -103.3 \\
9 & -102.7 & 47 & --- \\
\bottomrule
\end{tabularx}

\vspace{4pt}
{\footnotesize
\raggedright
$\lambda_i$ denotes the eigenvalue within the path DoF region $\Omega$.
\par
}
\label{tab:path_dof}
\end{table}

A UM-MIMO channel can be synthesized for DoF analysis with those estimated path parameters. On the basis of the UM-SIMO deployment, we further set the UM-MIMO geometry by extending the single antenna at Tx to a ULA array. Fig. \ref{measured scenario} illustrates the UM-MIMO setup, where the ULA has a length of 1 m and an element spacing of half a wavelength at 29 GHz (hence 193 elements in the ULA). Then, based on the estimated parameters $\{\hat{\alpha}, \hat{\tau}, \hat{\mathbf{\Theta}}\}$ for all paths, the UM-MIMO channel response can be generated via (\ref{ISP Gen}) and (\ref{ISP Gen1}).

Fig.~\ref{fig_second} and \ref{fig_third} illustrate the synthetic ULA-UCA channel after the reconstruction in the LoS scenario, and Fig. \ref{fig_second_nlos} and \ref{fig_third_nlos} show those in the NLoS scenarios. The CIRs across the UCA and ULA elements are extracted for demonstration. By comparing CIRs across the UCA in the measurements and reconstruction, i.e., Fig. \ref{fig_first} and \ref{fig_second}, and Fig. \ref{fig_first_nlos} and \ref{fig_second_nlos}, a good match between the measured and synthetic CIRs is observed, with all identified 9 paths in LoS and 5 paths in NLoS well reconstructed. Some minor mismatches may come from instability of the measurement platform during the measurement. Fig. \ref{fig_third} and \ref{fig_third_nlos} further illustrate the reconstruction responses from the Tx ULA side, where all detected paths in the measurements are clearly reconstructed. Note that their shapes are consistent with the propagation trajectories corresponding to the array geometry. These results demonstrate the effectiveness of the parameter estimation and channel reconstruction.

\subsection{DoF Result and Analysis}

\begin{figure}[t!]
\centering
\includegraphics[width=3.2in]{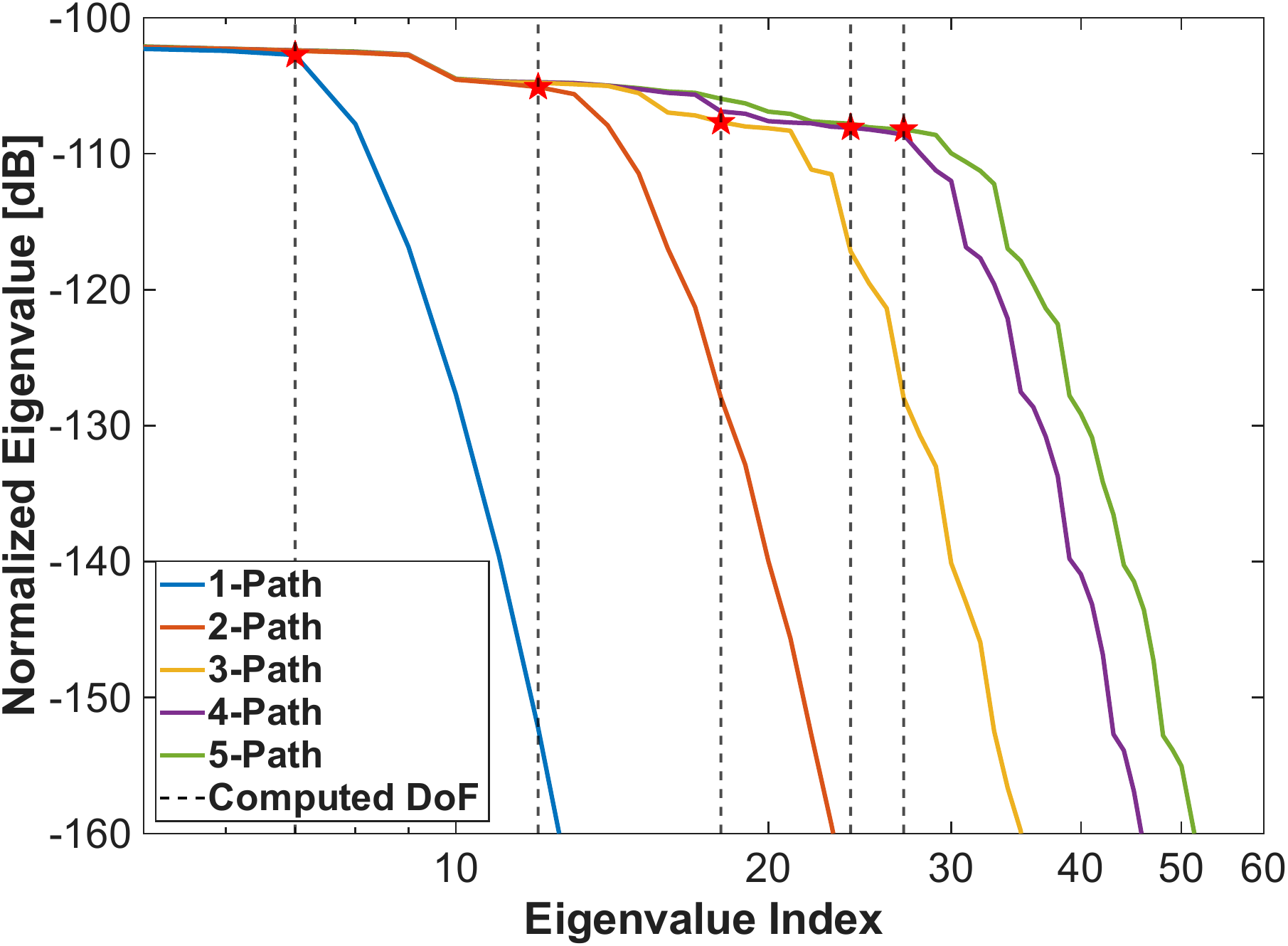}
\caption{Eigenvalue distribution of the synthetic ULA-UCA UM-MIMO NLoS channels, normalized by the array gains. The red stars mark the interaction points between the computed DoF values and eigenvalue distribution curves (with specific values listed in Table~\ref{tab:path_dof_nlos}).}
\label{final results_nlos}
\end{figure}

\begin{table}[t!]
\centering
\caption{Mapping relationship between eigenvalues and path power}
\renewcommand{\arraystretch}{1.2} 

\begin{tabularx}{\columnwidth}{c Y Y Y} 
\toprule
\textbf{Path $l$} & 
\textbf{Path power [dB]} & 
\textbf{Calculated DoF} &
\textbf{$\sum_{i\in \Omega_l} \lambda_i$ [dB]} \\ 
\midrule
1 & -92.6 & 7  & -92.6 \\ 
2 & -96.0 & 12 & -96.7 \\
3 & -97.9 & 18 & -97.5 \\
4 & -98.7 & 24 & -99.4 \\
5 & -102.1 & 27  & -103.4 \\ 

\bottomrule
\end{tabularx}
\label{tab:path_dof_nlos}
\end{table}

Once obtaining the UM-MIMO channel response, we perform the EVD to analyze the DoF result for validation of the proposed metric. Fig.~\ref{final results} illustrates the eigenvalue distributions in the LoS scenario, where channels with different numbers of superimposed paths are considered. Specifically, the ``$l$-path'' curve denotes the channel constructed using the first $l$ paths, where the path indices correspond to those in Fig.~\ref{subfig:LoS_meas_sce} and Fig.~\ref{fig_three_images}. As shown in Fig.~\ref{final results}, the computed DoF values via the proposed metric well match the target DoF values (i.e., the empirical rapid decay point in the distributions) for various cases (the minor deviations may come from the spatial correlation instability caused by the compact distribution of the UCA (compared to the ULA)). Particularly, for the LoS-transmission channel (i.e., ``1-path"), the proposed metric ccalculates the DoF value as 20, indicating an increase of DoF compared to the FF LoS case. Moreover, the channel DoF values increase as each effective path is added in the channel, and the computed DoF values consistently align with the decay points. However, the 4th, 5th, and 8th paths, corresponding to reflections from the back wall, ceiling, and ground, respectively, are not spatially resolvable from the LoS path by the ULA. As a result, the eigenvalue distributions overlap with their preceding curves, yielding no additional DoF gain. The proposed metric with the concept of effective summation in (\ref{eqn:doftotal}) captures this overlap phenomena. These results validate the effectiveness of the proposed metric criterion in predicting the DoF of multipath NF LoS channels. 

Table~\ref{tab:path_dof} presents the computed DoF and the summation of related eigenvalues. Two key findings are observed. First, the spatial DoF exhibits a cumulative growth as effective multipath components are included. Particularly, the DoF expands from 20 for the single LoS path to 48 for the complete channel (incorporating the 9 paths), representing an increase of more than 2 times. This substantial gain demonstrates that the spatial DoF contributed by multipaths is non-negligible, thereby indicating the significance of the proposed metric that can adequately consider multipath NF channels for UM-MIMO capacity analysis. Second, a quantitative correspondence is observed between the eigenvalue summation and the extracted path powers. Minor deviations may arise from path overlap and DoF prediction inaccuracies, which affect the eigenvalue summation range. This relationship provides insight for system performance analysis under different effective SNR conditions. For instance, when imposing a power threshold of 25 dB, the 8th and 9th paths can be neglected due to its weak power contribution. This implies that practical system designs may prioritize the spatial multiplexing gains of dominant paths while avoiding unnecessary resource allocation to weak sub-channels.

Fig.~\ref{final results_nlos} and Table~\ref{tab:path_dof_nlos} present the DoF analysis for the NLoS measurement scenario. The proposed metric accurately predicts the DoF values for channels with different numbers of paths, as indicated by the agreement between the computed DoF and the rapid decay points in the eigenvalue distributions. Specifically, the DoF increases from 7 for the single-path channel to 27 when all effective paths are considered, demonstrating the significant DoF contribution of multipath propagation. Moreover, the extracted path powers match well with the corresponding eigenvalue summations, further verifying the derived mapping relationship. Compared to the LoS scenario, the total DoF of the NLoS channel is smaller (27 versus 47), highlighting the importance of the LoS path for achieving higher spatial multiplexing capability. At the same time, since the channel DoF in NLoS scenarios relies entirely on multipath components, the necessity of the proposed metric becomes more evident, whereas existing LoS-based metrics cannot characterize such cases. These results validate the effectiveness of the proposed metric in capturing LoS-transmission, LoS multipath, and NLoS multipath channels. The proposed metric can be directly applied for practical DoF prediction, spatial multiplexing analysis, and capacity evaluation in NF UM-MIMO systems.

\section{Conclusion}
\label{sixth}

This paper investigates the spatial DoF of NF UM-MIMO channels under practical multipath propagation. Unlike existing studies limited to the NF LoS-transmission case, we derive a generic NF DoF metric considering both the LoS-transmission and multipath cases (including both LoS and NLoS). The DoF contributed by each path is determined by the product of the effective electrical aperture and the subtended solid angle, and the total channel DoF is obtained through the effective union of the spatially resolvable path contributions. In addition, a mapping between the eigenvalue distribution and the multipath powers is derived, providing a practical basis for identifying effective spatial modes under different power conditions.

Numerical simulations and real-world NF channel measurements at 28-30 GHz with a 720-element virtual array are conducted for validation, covering both LoS multipath and NLoS scenarios. The results show that the proposed metric accurately predicts the channel DoF, where the calculated values match the rapid decay points in the eigenvalue distributions. Specifically, the DoF increases from 20 in the LoS transmission to 47 in the measured LoS multipath scenario, and from 7 for the single-path channel to 27 when multiple paths are considered in the NLoS scenario. These results reveal the significant DoF contribution of multipath propagation and demonstrate that the proposed metric can effectively characterize NF channels beyond the conventional LoS-based analysis. The proposed metric therefore provides a practical tool for DoF prediction in NF UM-MIMO systems and can support future studies on capacity analysis, array deployment design, and spatial multiplexing optimization.

\bibliographystyle{IEEEtranchange}
\bibliography{reference}

\end{document}